\pdfoutput=1

\documentclass[conference]{IEEEtran}
\newcommand{\ignore}[1]{}
\usepackage[pass]{geometry}
\usepackage{fancyhdr}
\usepackage[normalem]{ulem}
\usepackage[hyphens]{url}
\usepackage{hyperref}
\usepackage{array}
\usepackage{color}
\usepackage{enumitem}
\usepackage{caption}
\usepackage{subcaption}
\usepackage{subfloat}
\usepackage{graphics}
\usepackage{amsmath}
\usepackage{comment}
\usepackage{nicefrac}
\usepackage{amssymb}
\usepackage{graphicx}
\usepackage{cuted}

\newtheorem{example}{Example}%[section]
\newtheorem{remark}{Remark}%[section]

\fancypagestyle{firstpage}{
  \fancyhf{}

  \pagenumbering{arabic}
}

\title{Achieving Multi-Port Memory Performance on Single-Port Memory with Coding Techniques}

\makeatletter
\newcommand{\linebreakand}{%
  \end{@IEEEauthorhalign}
  \hfill\mbox{}\par
  \mbox{}\hfill\begin{@IEEEauthorhalign}
}
\makeatother

\author{
\IEEEauthorblockN{Hardik Jain}
\IEEEauthorblockA{\textit{Department of ECE}\\
\textit{The University of Texas at Austin}\\
Austin, United States\\
hardikbjain@utexas.edu}
\and
\IEEEauthorblockN{Matthew Edwards\\}
\IEEEauthorblockA{\textit{GenXComm Inc.}\\
Austin, United States\\
matthewedwards@genxcomminc.com}
\and
\IEEEauthorblockN{Ethan Elenberg\\}
\IEEEauthorblockA{
\textit{ASAPP Inc.}\\
New York,$^{1}$ United States\\
eelenberg@asapp.com}
\linebreakand
\IEEEauthorblockN{Ankit Singh Rawat\\}
\IEEEauthorblockA{\textit{Google}\\
New York$^{1}$, United States\\
ankitsrwt@gmail.com}
\and
\IEEEauthorblockN{Sriram Vishwanath\\}
\IEEEauthorblockA{\textit{Department of ECE}\\
\textit{University of Texas at Austin}\\
Austin, United States\\
sriram@utexas.edu}

}

\begin{document}

\maketitle

\begin{abstract}
\footnotetext[1]{Work done primarily at The University of Texas at Austin}
Many performance critical systems today must rely on performance enhancements, such as multi-port memories, to keep up with the increasing demand of memory-access capacity. However, the large area footprints and complexity of existing multi-port memory designs limit their applicability. This paper explores a coding theoretic framework to address this problem. In particular, this paper introduces a framework to encode data across multiple single-port memory banks in order to {\em algorithmically} realize the functionality of multi-port memory.

This paper proposes three code designs with significantly less storage overhead compared to the existing replication based emulations of multi-port memories. To further improve performance, we also demonstrate a memory controller design that utilizes redundancy across coded memory banks to more efficiently schedule read and write requests sent across multiple cores. Furthermore, guided by DRAM traces, the paper explores {\em dynamic coding} techniques to improve the efficiency of the coding based memory design. We then show significant performance improvements in critical word read and write latency in the proposed coded-memory design when compared to a traditional uncoded-memory design.

\end{abstract}
\renewcommand\IEEEkeywordsname{Keywords}
%%%%%%%%%%%%%%%%%%%%%%%%%%%%%%%%%%%%%%%%%%%%%%%%%%%%%%%%
% Introduction
%%%%%%%%%%%%%%%%%%%%%%%%%%%%%%%%%%%%%%%%%%%%

\section{Introduction}
\label{sec:intro}

\begin{figure}[t!]
\centering
\includegraphics[width=0.6\linewidth]{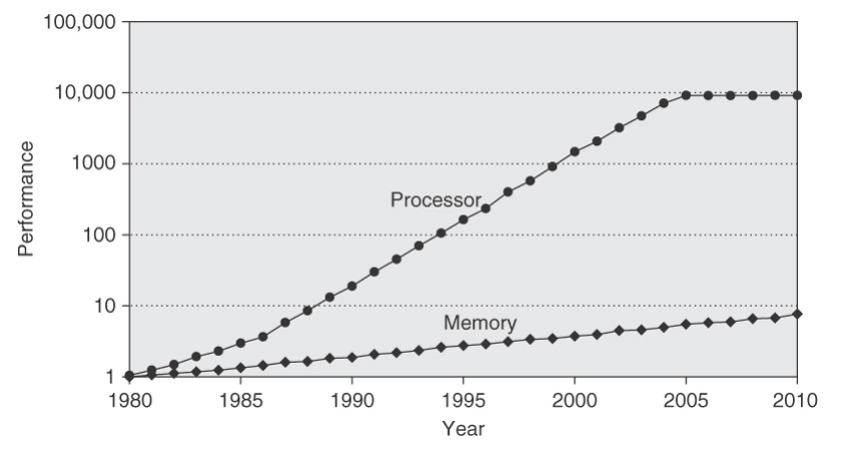}
\caption{\it{The gap in performance, measured as the difference in the time 
between processor memory requests for a single processor and the 
latency of a DRAM access~\cite{comparchbook}.}}
\label{fig:cpuvsmemory}
\end{figure}

Loading and storing information to memory is an intrinsic part of any computer program. As illustrated in Figure~\ref{fig:cpuvsmemory}, the past few decades have seen the performance gap between processors and memory grow. Even with the saturation and demise of Moore's law~\cite{Wulf1995, waldrop2016, MooreMITR}, processing power is expected to grow as multi-core architectures become more reliable~\cite{Geer}. The end-to-end performance of a program heavily depends on both processor and memory performance. Slower memory systems can bottleneck computational performance. This has motivated computer architects and researchers to explore strategies for shortening memory access latency, including sustained efforts towards enhancing the memory hierarchy~\cite{Burger}. Despite these efforts, long-latency memory accesses do occur when there is a miss in the last level cache (LLC). This triggers an access to shared memory, and the processor is stalled as it waits for the shared memory to return the requested information.

%---------------------------
\begin{figure}[t!]
\centering
\center{\includegraphics[width=.65\linewidth]{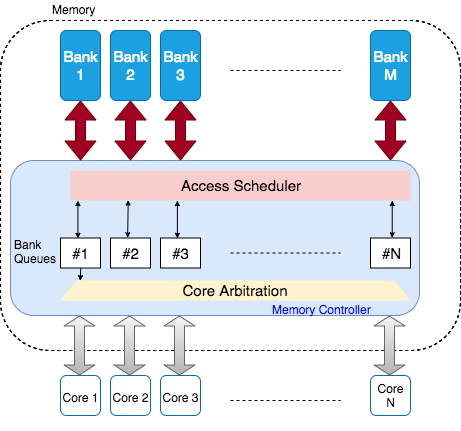}}
\caption{\it{General multi-core architecture with a shared memory. $N$ processor cores share a memory consisting of $M$ banks.}}
\label{fig:multicore_arch}
\end{figure}

%---------------------------
In multi-core systems, shared memory access conflicts between cores result in large access request queues. Figure~\ref{fig:multicore_arch}  illustrates a general multi-core architecture. The bank queues are served every memory clock cycle and the acknowledgement with data is sent back to the corresponding processor. In scenarios where multiple cores request access to memory locations in the same bank, the memory controller arbitrates them using bank queues. This contention between cores to access from the same bank is known as a {\em bank conflict}. As the number of bank conflicts increases, the resultant increases in memory access latency causes the multi-core system to slow.

%---------------------------
\begin{figure}[t!]
\centering
\includegraphics[width=0.3\linewidth]{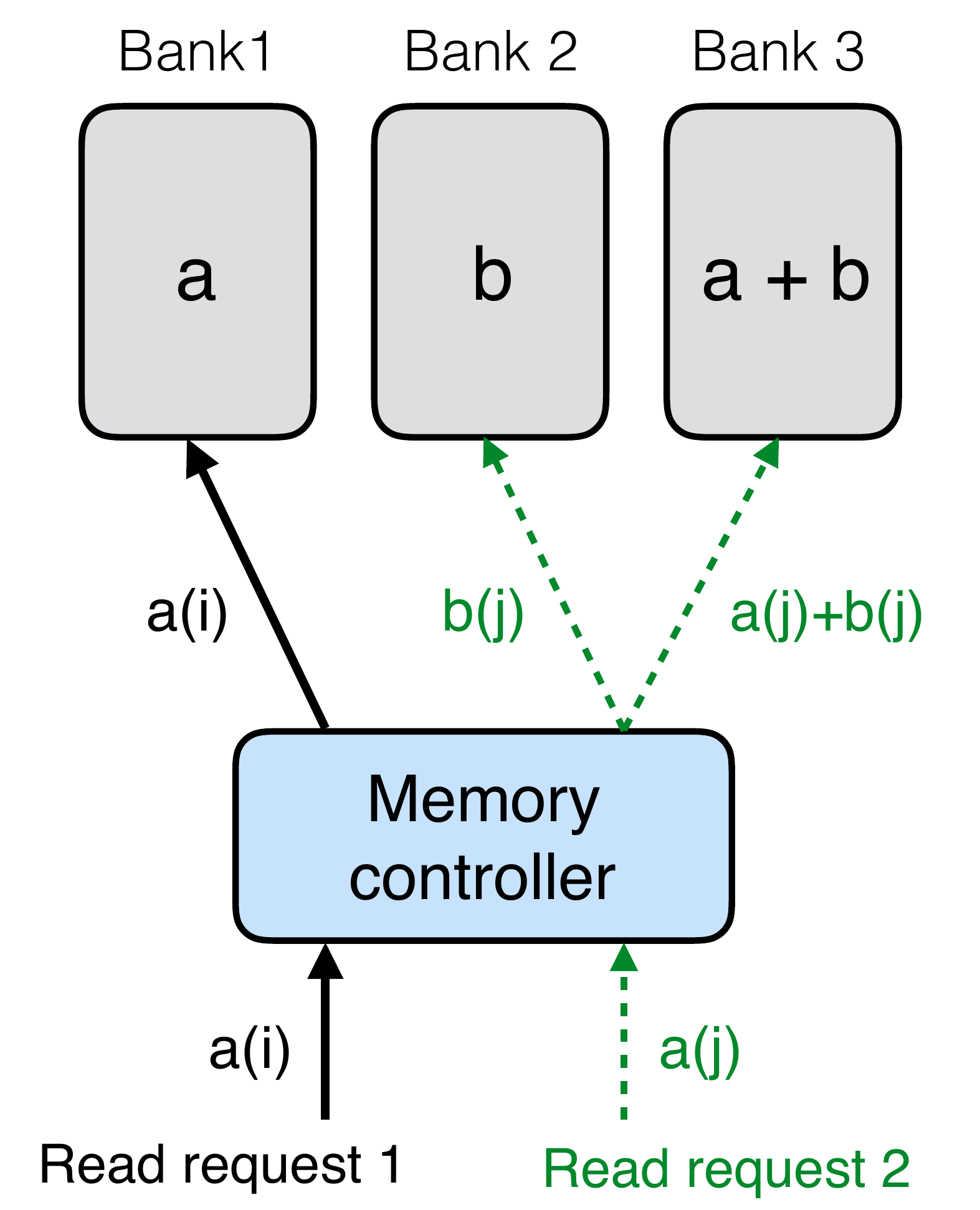}
\caption{\it{Here the redundant memory in Bank 3 enables multiple read accesses to Bank 1 or 2. Given two read requests $\{a(i), a(j)\}$ directed to Bank $1$, we can resolve bank conflict by reading $a(i)$ directly from Bank 1 and acquiring $a(j)$ with two reads from Bank 2 and Bank 3. $b(j)$ and $a(j) + b(j)$ are read from Bank 2 and Bank 3, and $a(j)$ is recovered because $a(j) = b(j) + a(j) + b(j)$.}}
\label{fig:example_xor}
\end{figure}

%---------------------------
We address the issue of increased latency by introducing a coded memory design. The main principle behind our memory design is to distribute accesses intended for a particular bank across multiple banks. We redundantly store encoded data, and we decode memory for highly requested memory banks using idle memory banks. This approach allows us to simultaneously serve multiple read requests intended for a particular bank. Figure~\ref{fig:example_xor} shows this with an example. Here, Bank 3 is redundant as its content is a function of the content stored on Banks 1 and 2. Such redundant banks are also referred to as {\em parity banks}. Assume that the information is arranged in $L$ rows in two first two banks, represented by $[a(1),\ldots, a(L)]$ and $[b(1),\ldots, b(L)]$, respectively. Let $+$ denote the XOR operation, and additionally assume that the memory controller is capable of performing simple decoding operations, \textit{i.e.} recovering $a(j)$ from $b(j)$ and $a(j) + b(j)$. Because the third bank stores $L$ rows containing $[a(1) + b(1),\ldots, a(L) + b(L)]$, this design allows us to simultaneously serve any two read requests in a single memory clock cycle.   

%---------------------------
%---------------------------
Hybrid memory designs such as the one in Figure~\ref{fig:example_xor} have additional requirements in addition to serving read requests. The presence of redundant parity banks raises a number of challenges while serving write requests. The memory overhead of redundant memory storage adds to the overall cost of such systems, so efforts must be made to minimize this overhead. Finally, the heavy memory access request rate possible in multi-core scenarios necessitates sophisticated scheduling strategies to be performed by the memory controller. In this paper we address these design challenges and evaluate potential solutions in a simulated memory environment. 

\noindent \textbf{Main contributions and organization:~}In this paper we systematically address all key issues pertaining to a shared memory system that can simultaneously service multiple access requests in a multi-core setup. We present all the necessary background on realization of multi-port memories using single-port memory banks along with an account of relevant prior work in Section~\ref{sec:bg}. We then present the main contributions of the paper which we summarize below.
\begin{itemize}
\item We focus on the design of the storage space in Section~\ref{sec:code_design}. In particular, we employ three specific coding schemes to redundantly store the information in memory banks. These coding schemes, which are based on the literature on distributed storage systems~\cite{dimakis, Gopalan12, batchcodes, RPDV16}, allow us to realize the functionality of multi-port memories from single port memories while efficiently utilizing the storage space.
\item We present a memory controller architecture for the proposed coding based memory system in Section~\ref{sec:memcontrol}. Among other issues, the memory controller design involves devising scheduling schemes for both read and write requests. This includes careful utilization of the redundancy present in the memory banks while maintaining the validity of information stored in them.
\item Focusing on applications where memory traces might exhibit favorable access patterns, we explore dynamic coding techniques which improve the efficiency of our coding based memory design in Sections~\ref{sec:dynamicCoding}.
\item Finally, we conduct a detailed evaluation of the proposed designs of shared memory systems in Section~\ref{sec:experimentalmethodology}. We implement our memory designs by extending Ramulator, a DRAM simulator~\cite{Ramulator}. We use the gem5 simulator~\cite{parsec_2_1_m5} to create memory traces of the PARSEC benchmarks~\cite{bienia09parsec2} which are input to our extended version of Ramulator. We then observe the execution-time speedups our memory designs yield.
\end{itemize}

%%%%%%%%%%%%%%%%%%%%%%%%%%%%%%%%%%%%%%%%%%%%%%%%%%%%%%%%
% Background and Related Work
%%%%%%%%%%%%%%%%%%%%%%%%%%%%%%%%%%%%%%%%%%%%
\section{Background and Related Work}
\label{sec:bg}

\subsection{Emulating multi-port memories}
\label{sec:emulation}

Multi-port memory systems are often considered to be essential for multi-core computation. Individual cores may request memory from the same bank simultaneously, and absent a multi-port memory system some cores will stall. Multi-port memory systems have significant design costs. Complex circuitry and area costs for multi-port bit-cells are significantly higher than those for single-port bit-cells~\cite{Suzuki,WLCH14}. This motivates the exploration of algorithmic and systematic designs that emulate multi-port memories using single-ported memory banks~\cite{ACP88, EMY91, RG91,Memoir_xor, Memoir_xor_virtual}. Attempts have been made to emulate multi-port memory using replication based designs \cite{CCES93}, however the resulting memory architectures are very large.

\subsubsection{Read-only Support} 
\label{sec:read_only}
\begin{figure}[t!]
\centering
\includegraphics[width=0.425\linewidth]{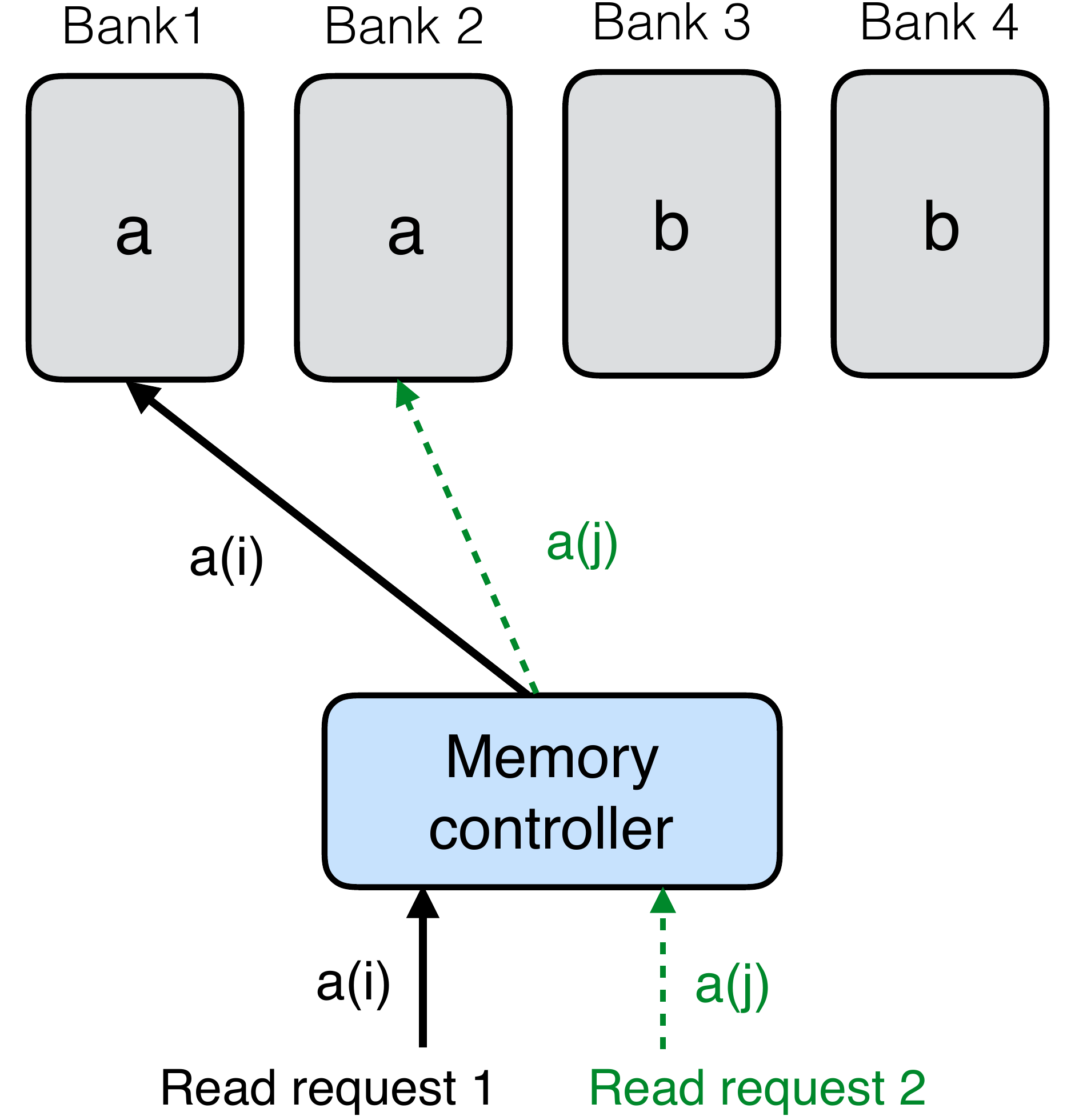}
\caption{\it{A $2$-replication design which supports $2$ read requests per bank. In this design, the data is partitioned between two banks $\mathbf{a} = [a(1),\ldots, a(L)]$ $\mathbf{b} = [b(1),\ldots, b(L)]$ and duplicated.}}
\label{fig:read_replication}
\end{figure}

Replication-based designs are often proposed as a method for multi-port emulation. Suppose that a memory design is required to support only read requests, say $r$ read requests per memory clock cycle. A simple solution is storing $r$ copies of each data element on $r$ different single-port memory banks. In every memory clock cycle, the $r$ read requests can be served in a straightforward manner by mapping all read request to distinct memory banks (see Figure~\ref{fig:read_replication}). This way, the $r$-replication design completely avoids bank conflicts for up to $r$ read request in a memory clock cycle. 

\begin{remark}
\label{rem:read_only}
If we compare the memory design in Figure~\ref{fig:read_replication} with that of Figure~\ref{fig:example_xor}, we notice that both designs can simultaneously serve $2$ read requests without causing any bank conflicts. Note that the design in Figure~\ref{fig:example_xor} consumes less storage space as it needs only $3$ single-port memory banks while the design in  Figure~\ref{fig:read_replication} requires $4$ single-port memory banks. However, the access process for the design in Figure~\ref{fig:example_xor} involves some computation. This observation raises the notion that sophisticated coding schemes allow for storage efficient designs compared to replication based methods~\cite{MacSlo}. However, this comes at the expense of increased computation required for decoding.
\end{remark}

%---------------------------
%---------------------------

%---------------------------
\begin{figure}[t!]
\centering
\includegraphics[width=0.60\linewidth]{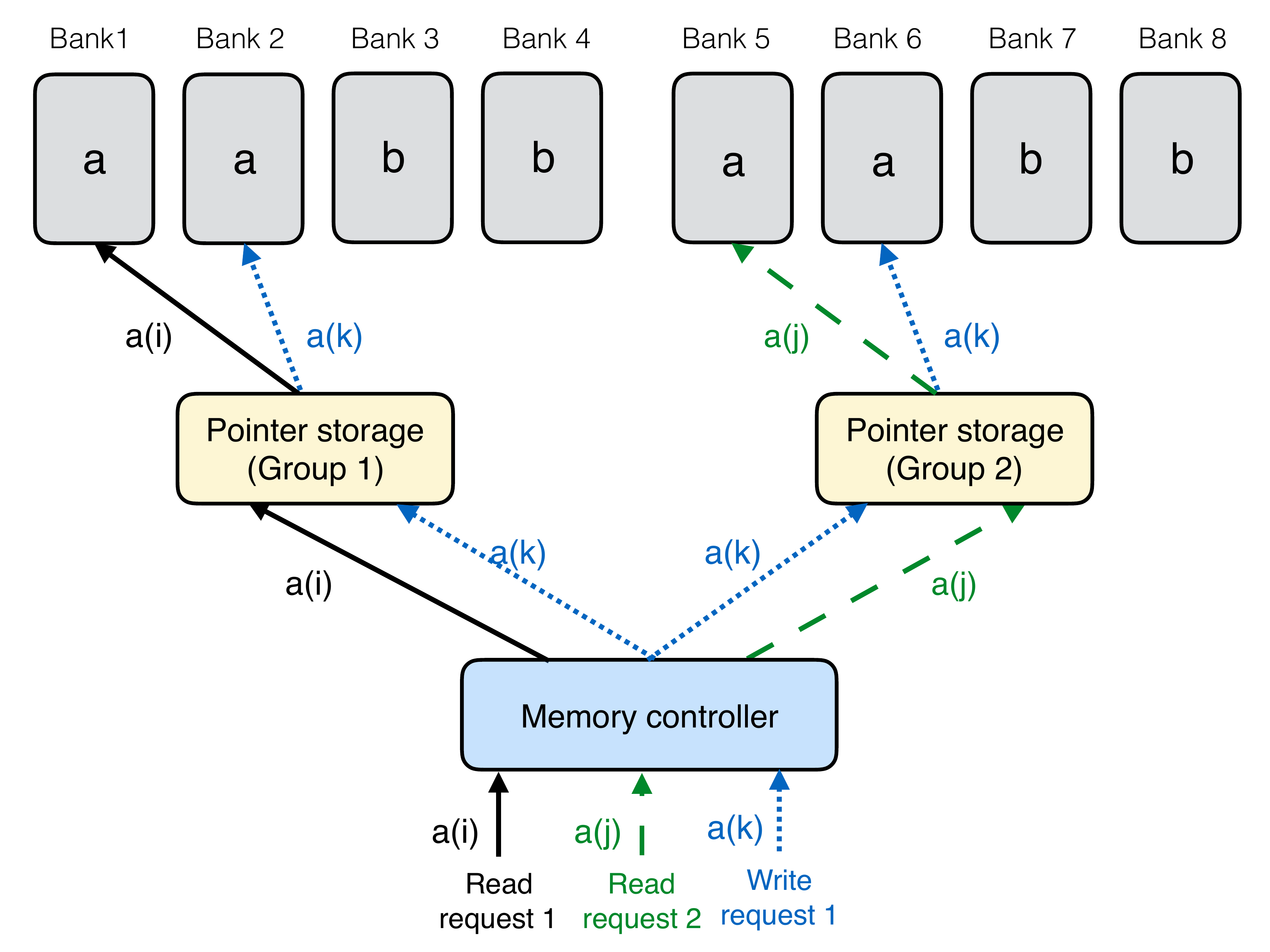}
\caption{\it{A $4$-replication based design to support $r = 2$ read requests and $w = 1$ write requests. Both collections of information elements $\mathbf{a} = [a(1),\ldots, a(L)]$ and $\mathbf{b} = [b(1),\ldots, b(L)]$ are replicated to obtain $r\cdot (w + 1) = 4$ single-port memory banks. These banks are then partitioned into $r = 2$ disjoint groups, Banks $1$ -- $4$ and Banks $5$ -- $8$. The pointer storage is required to ensure all read requests are not served stale symbols. As shown in the illustration, the write request is served to two of the $\mathbf{a}$ banks to ensure that the fresh $a(k)$ may be served during any future cycle.}}

\label{fig:rw_replication}
\end{figure}
%---------------------------
\subsubsection{Read and Write Support}
\label{sec:rw}
A proper emulation of multi-port memory must be able to serve  write requests. A challenge that arises from this requirement is tracking the state of memory. In replication-based designs where original data banks are duplicated, the service of writes requests results in differences in state between the original and duplicate banks.

Replication-based solutions to the problems presented when supporting write requests involve creating yet more duplicate banks. A replication-based multi-port memory emulation that simultaneously supports $r$ read requests and $w$ write requests requires a $r\cdot(w + 1)$ replication scheme, where $r\cdot(w+1)$ copies of each data element are stored on $r\cdot(w + 1)$ different single-port memory banks. We illustrate this scheme for $r = 2$ and $w = 1$ in Figure~\ref{fig:rw_replication}. As in previous illustrations, we have two groups of symbols $\mathbf{a} = [a(1),\ldots, a(L)]$ and $\mathbf{b}  = [b(1),\ldots, b(L)]$. We store $4$ copies each of data elements $\mathbf{a}$ and $\mathbf{b}$ and partition the banks into $r = 2$ disjoint groups. Each group contains $(w + 1) = 2$ memory banks. An additional storage space, the pointer storage, is required to keep track the state of the data in the banks.

\subsection{Storage-efficient emulation of multi-port memories}
\label{sec:efficient_emulation}

As described in Section~\ref{sec:emulation}, introducing redundancy to systems which use single-port memory banks allows such systems to emulate the behavior of multi-port banks. Emulating multi-port read and write systems is costly (cf. Section~\ref{sec:rw}). A greater number of single-port memory banks are needed, and systems which redundantly store memory require tracking of the various versions of the data elements present in the memory banks. Furthermore, as write requests are served the elements stored across redundant banks temporary differ. This transient inconstancy between redundant storage complicates the process of arbitration.

We believe that various tasks that arise in the presence of write requests and contribute to computational overhead of the memory design, including synchronization among memory banks and complicated arbitration, can be better managed at the algorithmic level. Note that these tasks are performed by the memory controller. It is possible to mitigate the effect of these tasks on the memory system by relying on the increasing available computational resources while designing the memory controller. Additionally, we believe that large storage overhead is a more fundamental issue that needs to be addressed before multi-port memory emulation is feasible. In particular, the large replication factor in a naive emulation creates such a large storage overhead that the resulting area requirements of such designs are impractical.

Another approach arises from the observation that some data banks are left unused during arbitration in individual memory cycles, while other data banks receive multiple requests. We encode the elements of the data banks using specific coding schemes to generate parity banks. Elements drawn from multiple data banks are encoded and stored in the parity banks. This approach allows us to utilize idle data banks to decode elements stored in the parity banks in service of multiple requests which target the same data bank. We recognize that this approach leads to increased complexity at the memory controller. However, we show that the increase in complexity can be kept within an acceptable level while ensuring storage-efficient emulation of multi-port memories.

\subsection{Related work}

Coding theory is a well-studied field which aims to mitigate the challenges of underlying mediums in information processing systems ~\cite{MacSlo, Cover}. The field has enabled both reliable communication across noisy channels and reliability in fault-prone storage units. Recently, we have witnessed intensive efforts towards the application of coding theoretic ideas to design large scale distributed storage systems \cite{Azure, SAPDVCB13, Rashmi14}. In this domain, the issue of access efficiency has also received attention, especially the ability to support multiple simultaneous read accesses with small storage overhead~\cite{batchcodes, RPDV16, RSDG16, Wang2017}. In this paper, we rely on such coding techniques to emulate multi-port memories using single-port memory banks. We note that the existing work on batch codes~\cite{batchcodes} focuses only on read requests, but the emulation of multi-port memory must also handle write requests. 

Coding schemes with low update complexity that can be implemented at the speed memory systems require have also been studied ~\cite{ASV10, MCW14}. Our work is distinguished from the majority of the literature on coding for distributed storage, because we consider the interplay between read and write requests and how this interplay affects memory access latency.

The work which is closest to our solution for emulating a multi-port memory is by Iyer and Chuang~\cite{Memoir_xor, Memoir_xor_virtual}, where they also employ XOR based coding schemes to redundantly store information in an array of single-port memory banks. However, we note that our work significantly differers from \cite{Memoir_xor, Memoir_xor_virtual} as we specifically rely on different coding schemes arising under the framework of batch codes~\cite{batchcodes}. Additionally, due to the employment of distinct coding techniques, the design of memory controller in our work also differs from that in \cite{Memoir_xor, Memoir_xor_virtual}.

%%%%%%%%%%%%%%%%%%%%%%%%%%%%%%%%%%%%%%%%%%%%%%%%%%%%%%%%
% Code design
%%%%%%%%%%%%%%%%%%%%%%%%%%%%%%%%%%%%%%%%%%%%
\section{Codes to Improve Accesses}
\label{sec:code_design}

Introducing redundancy into a storage space comprised of single-port memory banks enables simultaneous memory access. In this section we propose memory designs that utilize coding schemes which are designed for access-efficiency. We first define some basic concepts with an illustrative example and then describe $3$ coding schemes in detail.

\subsection{Coding for memory banks}
\label{sec:coding_mb}

A coding scheme defines how memory is encoded to yield redundant storage. The memory structures which store the original memory elements are known as {\em data banks}. The elements of the data banks go through an {\em encoding process} which generates a number of {\em parity banks}.  The parity banks contain elements constructed from elements drawn from two or more data banks. A linear encoding process such as XOR may be used to minimize computational complexity. The following example further clarifies these concepts and provides some necessary notation.

\begin{figure}[t!]
\centering
  \includegraphics[width=0.45\linewidth]{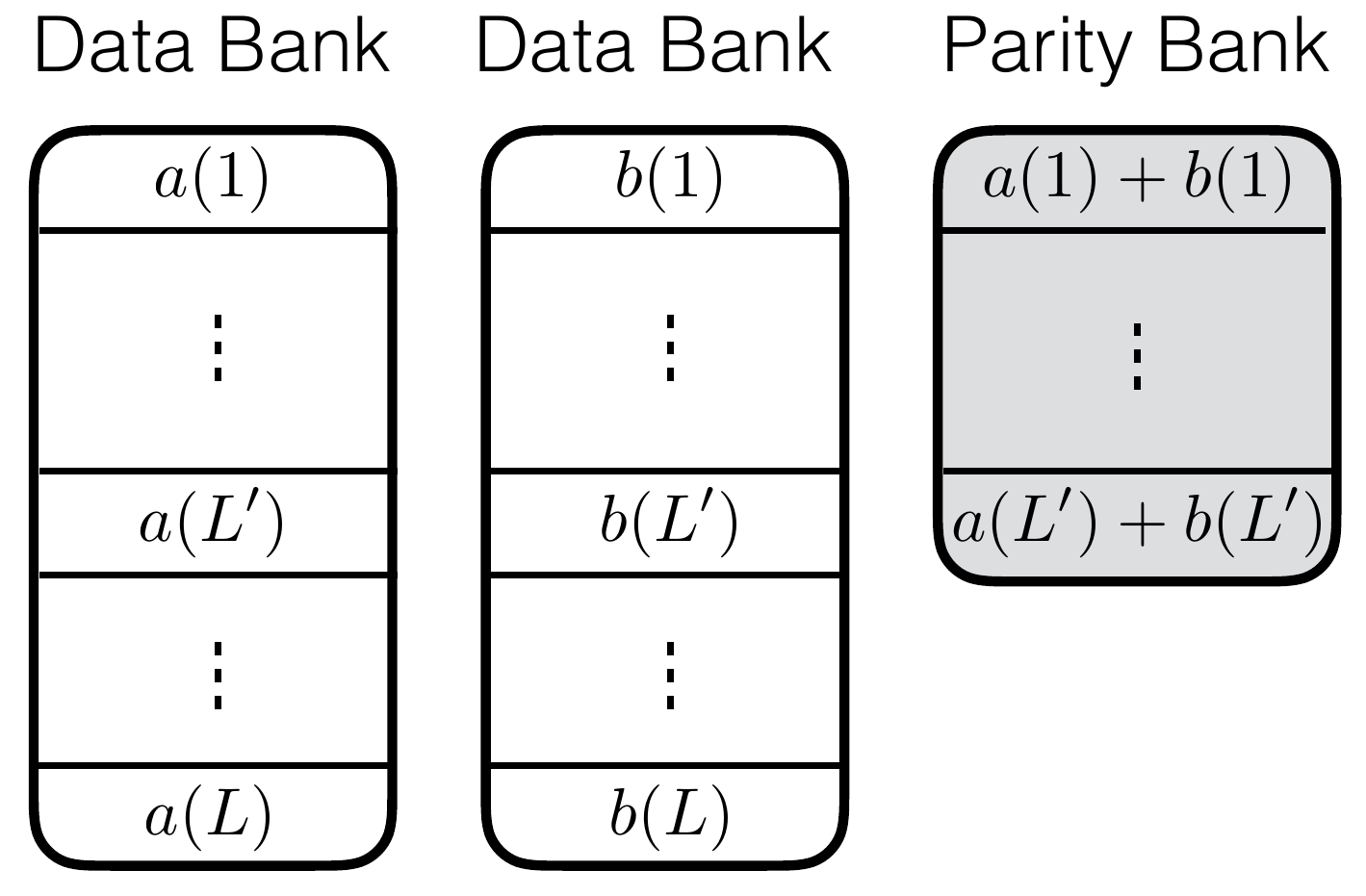} 
\caption{\it{This illustration is an example parity design.}}
\label{fig:example1}
\end{figure}

\begin{example}
Consider a setup with two data banks $\mathbf{a}$ and $\mathbf{b}$. We assume that each of the banks store $L \cdot W$ binary data elements\footnote{It is possible to work with data elements over larger alphabets/finite fields. However, assuming data elements to be binary suffices for this paper as only work with coding schemes defined over binary field.} which are arranged in an $L \times W$ array. In particular, for $i \in [L] \triangleq \{1,\ldots, L\}$, $a(i)$ and $b(i)$ denote the $i$-th row of the bank $\mathbf{a}$ and bank $\mathbf{b}$, respectively. Moreover, for $i \in [L]$ and $j \in [W] \triangleq \{1,\ldots, W\}$, we use $a_{i, j}$ and $b_{i, j}$ to denote the $j$-th element in the rows $a(i)$ and $b(i)$, respectively. Therefore, for $i \in [L]$, we have 
\begin{align}
a(i) = \big(a_{i,1}, a_{i,2},\ldots, a_{i, W}\big) \in \{0, 1\}^W\nonumber \\
b(i) = \big(b_{i,1}, b_{i,2},\ldots, b_{i, W}\big) \in \{0, 1\}^W. \nonumber
\end{align}
Now, consider a linear coding scheme that produces a parity bank $\mathbf{p}$ with $L'W$ bits arranged in an $L' \times W$ array such that for $i \in [L'] \triangleq \{1,\ldots, L'\}$, 
\begin{align}
p(i) &= \big(p_{i, 1},\ldots, p_{i,W}\big) = a(i) + b(i) \nonumber \\
&\triangleq \left(a_{i,1} + b_{i,1}, a_{i,1} + b_{i,1},\ldots, a_{i,1} + b_{i,1}\right). 
\end{align}
\end{example}
\begin{remark}
Figure~\ref{fig:example1} illustrates this coding scheme. Because the parity bank is based on those rows of the data banks that are indexed by the set $[L'] \subseteq [L]$, we use the following concise notation to represent the encoding of the parity bank. 
$$
\mathbf{p} = \mathbf{a}([L']) +  \mathbf{b}([L']).
$$
In general, we can use any subset $\mathcal{S} = \{i_1, i_2,\ldots, i_{L'}\} \subseteq [L]$ comprising $L'$ rows of data banks to generate the parity bank $\mathbf{p}$. In this case, we have $\mathbf{p} = \mathbf{a}(\mathcal{S}) +  \mathbf{b}(\mathcal{S})$, or
\begin{align*}
p(l) = a(i_l) + b(i_l)~\text{for}~l \in [L'].
\end{align*}
\end{remark}

\begin{remark}
Note that we allow for the data banks and parity banks to have different sizes, \textit{i.e.} $L \neq L'$. This freedom in memory design can be utilized to reduce the storage overhead of parity banks based on the underlying application. If the size of a parity bank is smaller than a data bank, \textit{i.e.} $L' < L$, we say that the parity bank is a {\em shallow bank}. We note that it is reasonable to assume the existence of shallow banks, especially in proprietary designs of integrated memories in a system on a chip (SoC).
\end{remark}

\begin{remark}
\label{rem:design1}
Note that the size of shallow banks is a design choice which is controlled by the parameter $0 < \alpha \leq 1$. A small value of $\alpha$ corresponds to small storage overhead. The choice of a small $\alpha$ comes at the cost of limiting parity memory accesses to certain memory ranges. In Section~\ref{sec:dynamicCoding} we discuss techniques for choosing which regions of memory to encode. In scenarios where many memory accesses are localized to small regions of memory, shallow banks can support many parallel memory accesses for little storage overhead. For applications where memory access patterns are less concentrated, the robustness of the parity banks allows one to employ a design with $\alpha = 1$.
\end{remark}

\subsubsection{Degraded reads and their locality}
\label{sec:degraded}

The redundant data generated by a coding scheme mitigates bank conflicts by supporting multiple read accesses to the original data elements. Consider the coding scheme illustrated in Figure~\ref{fig:example1} with a parity bank $\mathbf{p} = \mathbf{a}([L']) + \mathbf{b}([L'])$. In an uncoded memory system simultaneous read requests for bank $\mathbf{a}$, such as $a(1)$ and $a(5)$, result in a bank conflict. The introduction of $\mathbf{p}$ allows both read requests to be served. First, $a(1)$ is served directly from bank  $\mathbf{a}$. Next, $b(5)$ and $p(5)$ are downloaded. $a(5) = b(5) + p(5)$, so $a(5)$ is recovered by means of the memory in the parity bank. A read request which is served with the help of parity banks is called a {\em degraded read}. Each degraded read has a parameter {\em locality} which corresponds to the total number of banks used to serve it. Here, the degraded read for $a(5)$ using $\mathbf{b}$ and $\mathbf{p}$ has locality $2$.
\subsection{Codes to emulate multi-port memory}
\label{sec:designs}

We will now describe the code schemes proposed for the emulation of multi-port memories. Among a large set of possible coding schemes, we focus on three specific coding schemes for this task. We believe that these three coding schemes strike a good balance among various quantitative parameters, including storage overhead, number of simultaneous read requests supported by the array of banks, and the locality associated with various degraded reads. Furthermore, these coding schemes respect the practical constraint of encoding across a small number of data banks. In particular, we focus on the setup with $8$ memory banks.
\subsubsection{Code Scheme I}
\label{sec:design1}

\begin{figure}[ht!]
\centering
	\center{\includegraphics[width=.85\linewidth]{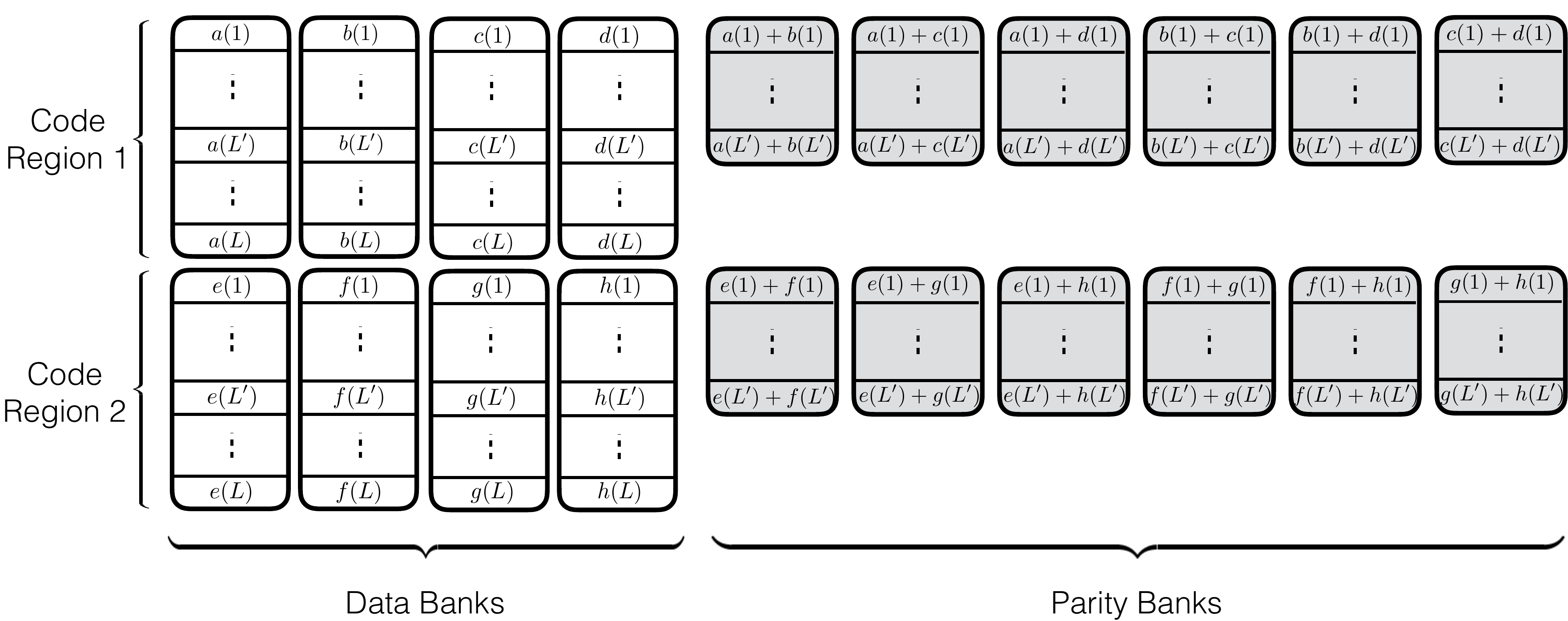}}
	\caption{\it{Pictured here is an illustration of code scheme I.}}
	\label{fig:design1}
\end{figure}

This code scheme is motivated from the concept of batch codes~\cite{batchcodes} which enables parallel access to content stored in a large scale distributed storage system.
The code scheme involves $8$ data banks $\{\mathbf{a}, \mathbf{b},\ldots, \mathbf{h}\}$ each of size $L$ and $12$ shallow banks each of size $L' = \alpha L$. We partition the $8$ data banks into two  groups of $4$ banks. The underlying coding scheme produces shallow parity banks by separately encoding data banks from the two groups. Figure~\ref{fig:design1} shows the resulting memory banks. The storage overhead of this schemes is $12\alpha L$ which implies the rate\footnote{The information rate is a standard measure of redundancy of a coding scheme ranging from $0$ to $1$, where $1$ corresponds to the most efficient utilization of storage space.} of the coding scheme is $$\frac{8L}{8L + 12\alpha L} = \frac{2}{2 + 3\alpha}.$$

We now analyze the number of simultaneous read requests that can be supported by this code scheme. \\

\noindent \textbf{Best case analysis:~}This code scheme achieves maximum 
performance when sequential accesses to the coded regions are issued. During the 
best case access, we can achieve up to $10$ parallel accesses to a particular coded region in one access cycle.
Consider the scenario when we receive accesses to the following $10$ rows:
\begin{align*}
&\left\{a(1),b(1),c(1),d(1),a(2),b(2),c(2),d(2),c(3),d(3)\right\} .
\end{align*}
Note that we can serve the read requests for the rows \\ $\{a(1),b(1),c(1),d(1)\}$ using the data bank $\mathbf{a}$ and the three parity banks storing $\{a(1)+b(1), b(1)+c(1),c(1)+d(1)\}$. The requests for $\{a(2),c(2),d(2)\}$ can be served by downloading $b(2)$ from the data bank $\mathbf{b}$ and $\{a(2)+d(2), b(2)+d(2),a(2)+c(2)\}$ from their respective parity banks. Lastly, in the same memory clock cycle, we can serve the requests for $\{c(3), d(3)\}$ using the data banks $\mathbf{c}$ and $\mathbf{d}$.\\
%------------------------------
%------------------------------
\ignore{
%------------------------------
\begin{figure}[ht!]
\centering
\includegraphics[width=150mm,natwidth=610,natheight=642]{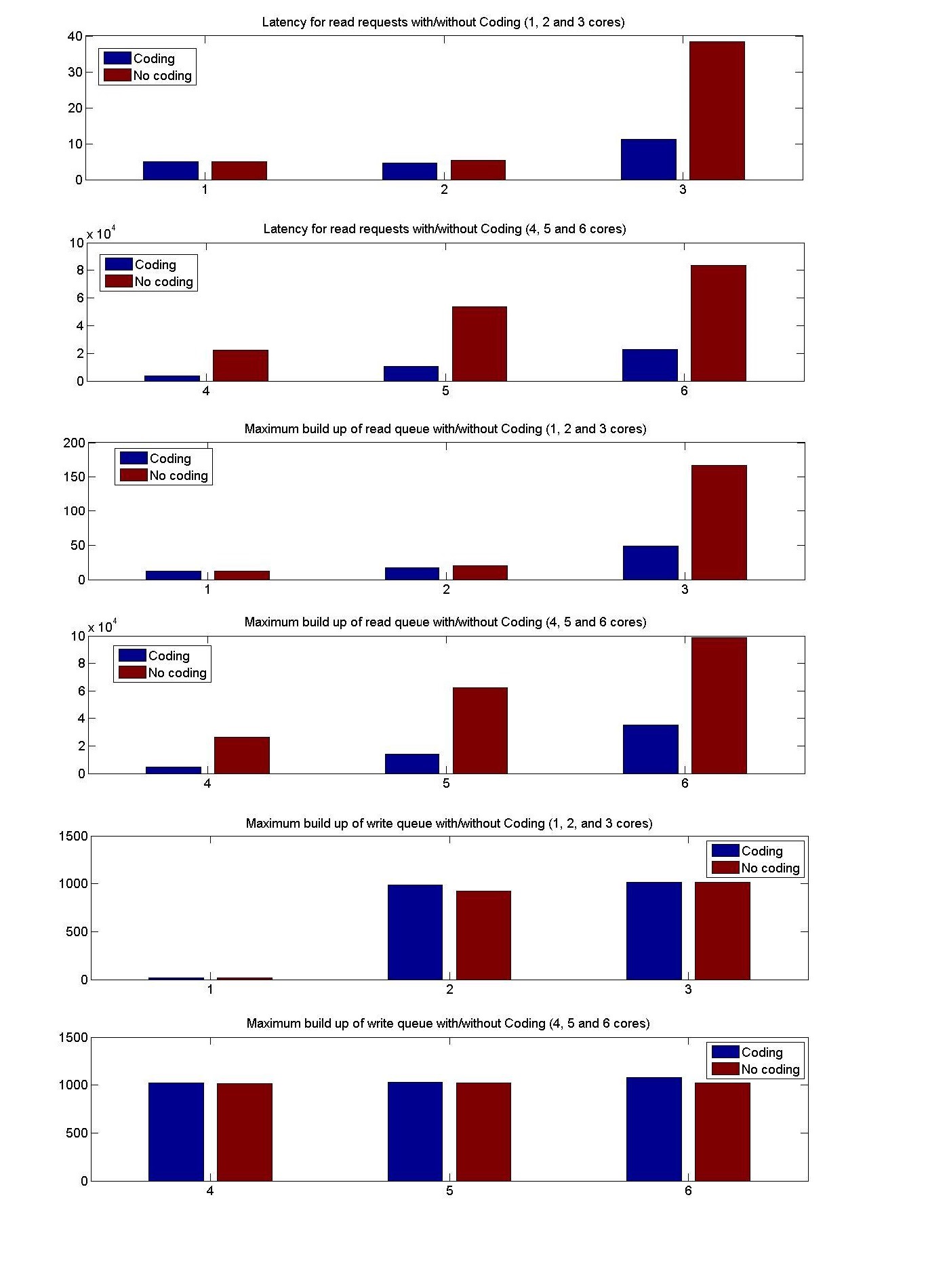}
\caption{ }
\label{fig:result_design1}
\end{figure}
%------------------------------
}
\noindent \textbf{Worst case analysis}: This code scheme  (cf.~Figure~\ref{fig:design1}) may fail to utilize any parity banks depending on the requests waiting to be served. The worst case scenario for this code scheme is when there are non-sequential and non-consecutive access to the memory 
banks. Take for example a scenario where we only consider the first four banks of the code scheme. The following read requests are waiting to be served:  
\begin{align*}
\{a(1), a(2), b(8), b(9), c(10),c(11), d(14), d(15)\}. 
\end{align*}
Because none of the requests share the same row index, we are unable to utilize the parity banks. The worst case number of reads per cycle is equal to the number of data banks. 

\subsubsection{Code Scheme II}
\label{sec:design2}

\begin{figure}[!ht]
	\center{\includegraphics[width=.85\linewidth]{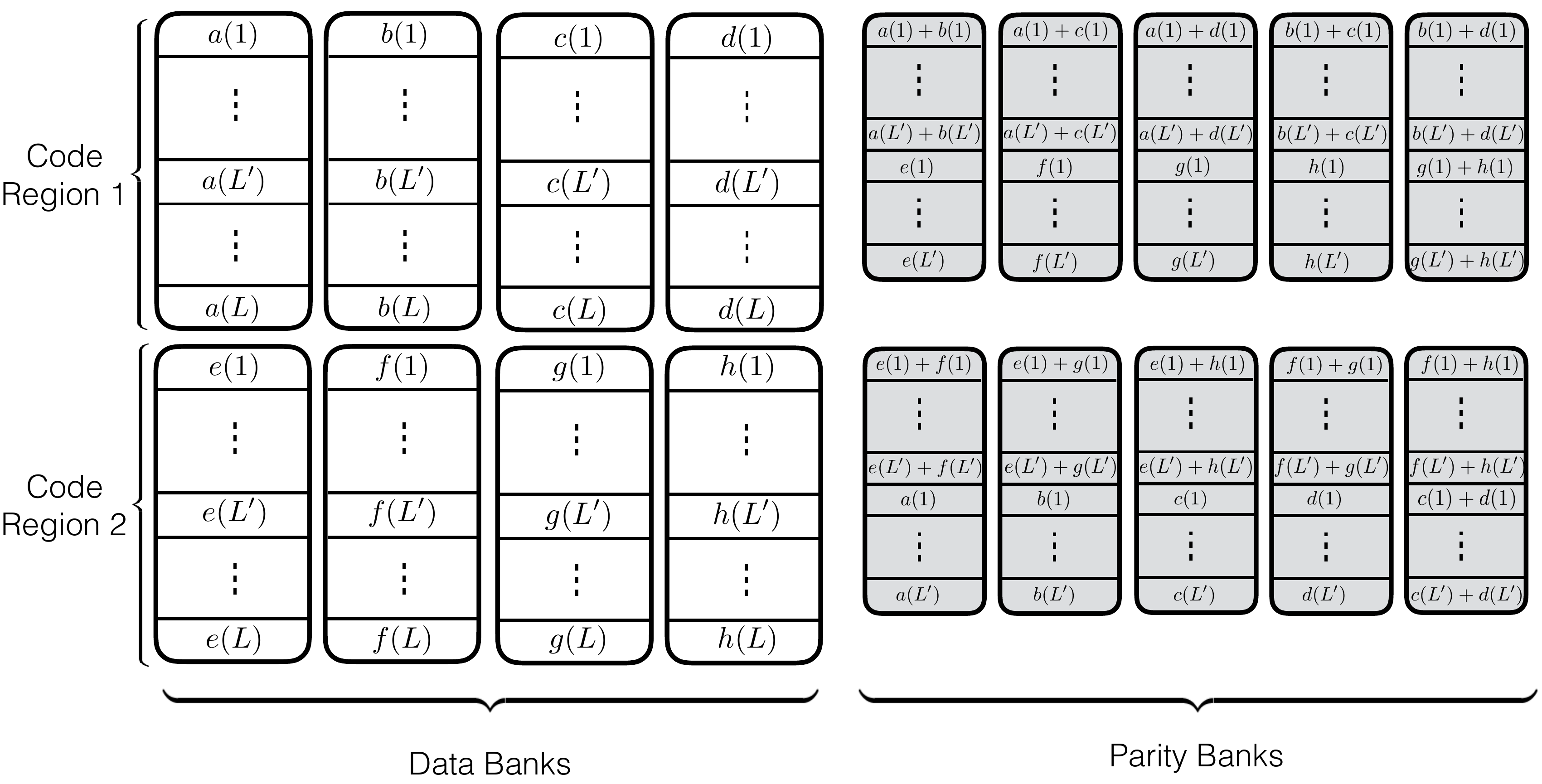}}
	\caption{\it{Pictured here is an illustration of code scheme II.}}
	\label{fig:design2}
\end{figure}

Figure~\ref{fig:design2} illustrates the second code scheme explored in this paper. Again, the $8$ data banks $\{\mathbf{a}, \mathbf{b},\ldots, \mathbf{h}\}$ are partitioned into two groups containing $4$ data banks each. These two groups are then associated with two code regions. The first code region is similar to the previous code scheme, as it contains parity elements constructed from two data banks. The second code region contains data directly duplicated from single data banks. This code scheme further differs from the previous code scheme (cf. Figure~\ref{fig:design1}) in terms of the size and arrangement parity banks. Even though $L' = \alpha L$ rows from each data bank are stored in a coded manner by generating parity elements, the parity banks are assumed to be storing $2\alpha L > L'$ rows.

For a specific choice of $\alpha$, the storage overhead of this scheme is $20\alpha L$ which leads to a rate of $$\frac{8L}{8L + 20\alpha L} = \frac{2}{2 + 5\alpha}.$$ Note that this code scheme can support $5$ read accesses per data bank in a single memory clock cycle as opposed to $4$ read requests supported by the code scheme from Section~\ref{sec:design1}. However, this is made possible at the cost of extra storage overhead. Next, we discuss the performance of this code scheme in terms of the number of simultaneous read requests that can be served in the best and worst case.

\noindent \textbf{Best case analysis:~} This code scheme achieves the best access performance when sequential accesses to the data banks are issued. In particular, this scheme can support up to $9$ read requests in a single memory clock cycle. Consider the scenario where we receive read requests for the following rows of the data banks:
$$
\big\{a(1),b(1),c(1),d(1),a(2),b(2),c(2),d(2),a(3),b(3),c(3)\big\}.
$$ Here, we can serve 
$\{a(1), b(1), c(1), d(1)\}$ using the data bank $\mathbf{a}$ with the parity banks storing the parity elements $\{a(1) + b(1),b(1)+c(1),c(1)+d(1)\}$. Similarly, we can serve the requests for the rows $\{a(2),b(2),d(2)\}$ using the data bank $\mathbf{b}$ with the parity banks storing the parity elements $\{a(2)+d(2), b(2)+d(2)\}$. Lastly, the request for the rows $c(2)$ and $d(3)$ is served using the data banks $\mathbf{c}$ and $\mathbf{d}$.\\

\noindent \textbf{Worst case analysis:~}Similar to the worst case in Scheme I, this code scheme can enable $5$ simultaneous accesses in a single memory clock cycle in the
worst case. The worst case occurs when requests are non-sequential and non-consecutive.

\subsubsection{Code Scheme III}
\begin{figure}[!ht]
	\centering
	\begin{minipage}[!t]{\linewidth}
		\center{\includegraphics[width=.85\linewidth]{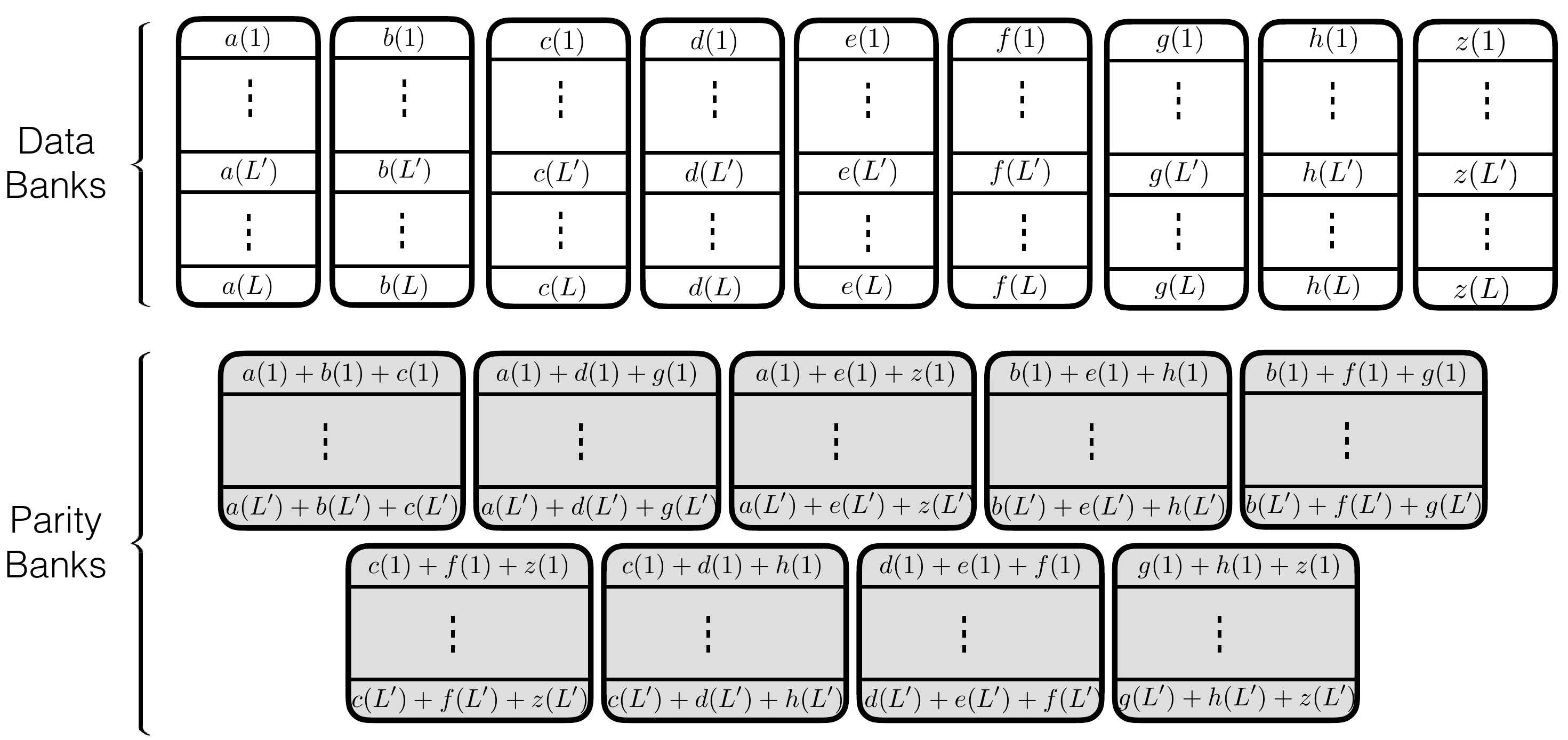}}
		\caption{\it{Pictured here is an illustration of code scheme III.}}
		\label{fig:design3}
	\end{minipage}
\end{figure}

The next code scheme we discuss has locality 3, so each degraded read requires two parity banks to be served. This code scheme works with $9$ data bank $\{\mathbf{a}, \mathbf{b},\ldots, \mathbf{h}, \mathbf{z}\}$ and generates $9$ shallow parity banks. Figure~\ref{fig:design3} shows this scheme.
The storage overhead of this scheme is $9\alpha L$ which corresponds to the rate of $\frac{1}{1 + \alpha}$. We note that this scheme possesses higher logical complexity as a result of its increased locality. 

This scheme supports $4$ simultaneous read access per bank per memory clock cycle as demonstrated by the following example. Suppose rows $\{a(1), a(2), a(3), a(4)\}$ are requested. $a(1)$ can be served directly from $\mathbf{a}$. $a(2)$ is served by means of a parity read and reads to banks $\mathbf{b}$ and $\mathbf{c}$, $a(3)$ is served by means of a parity read and reads to banks $\mathbf{d}$ and $\mathbf{g}$, and $a(4)$ is served by means of a parity read and reads to banks $\mathbf{e}$ and $\mathbf{z}$.

\noindent \textbf{Best case analysis:~} Following the analysis similar to code schemes I and II, the best case number of reads per cycle will be equal to the number of data and parity banks.

\noindent \textbf{Worst case analysis:~} Similar to code schemes I and II, the number of reads per cycle is equal to the number of data banks. 

%---------------------------------------

\begin{remark}
Note that the coding scheme in Figure~\ref{fig:design3} describes a system with $9$ data banks. However, we have set out to construct a memory system with $8$ data banks. It is straightforward to modify this code scheme to work with $8$ data banks by simple omitting the final data bank from the encoding operation.
 \end{remark}

%%%%%%%%%%%%%%%%%%%%%%%%%%%%%%%%%%%%%%%%%%%%%%%%%%%%%%%%
% Memory controller design
%%%%%%%%%%%%%%%%%%%%%%%%%%%%%%%%%%%%%%%%%%%%
\section{Memory Controller Design}
\label{sec:memcontrol}
The architecture of the memory controller is focused on exploiting redundant storage in the coding schemes to serve memory requests faster than an uncoded scheme.

The following three stages are illustrated in Figure~\ref{fig:multicore_arch}:
\begin{itemize}
\item \textbf{Core arbiter:~}Every clock cycle, the \textit{core arbiter} receives up to one request from each core which it stores in an internal queue. The core arbiter attempts to push these requests to the appropriate bank queue. If it detects that the destination bank queue is full, the controller signals that the core is busy which stalls the core.

\item \textbf{Bank queues:~}Each data bank has a corresponding \textit{read queue} and \textit{write queue}.  The core arbiter sends memory requests to the bank queues until the queues are full. In our simulations, we use a bank queue depth of 10. There is also an additional queue which holds special requests such as memory refresh requests.

\item \textbf{Access scheduler:~}Every memory cycle, the \textit{access scheduler} chooses to serve read requests or write requests, algorithmically determining which requests in the bank queues it will schedule. The scheduling algorithms the access scheduler uses are called pattern builders. Depending on the current memory cycle's request type, the access scheduler invokes either the read or write pattern builder.
\end{itemize}

We note that the core arbiter and bank queues should not differ much from those in a traditional uncoded memory controller. The access scheduler directly interacts with the memory banks, and therefore must be designed specifically for our coding schemes.

%-----------------------
\begin{figure}[tbp]
\centering
\includegraphics[width=0.6\linewidth]{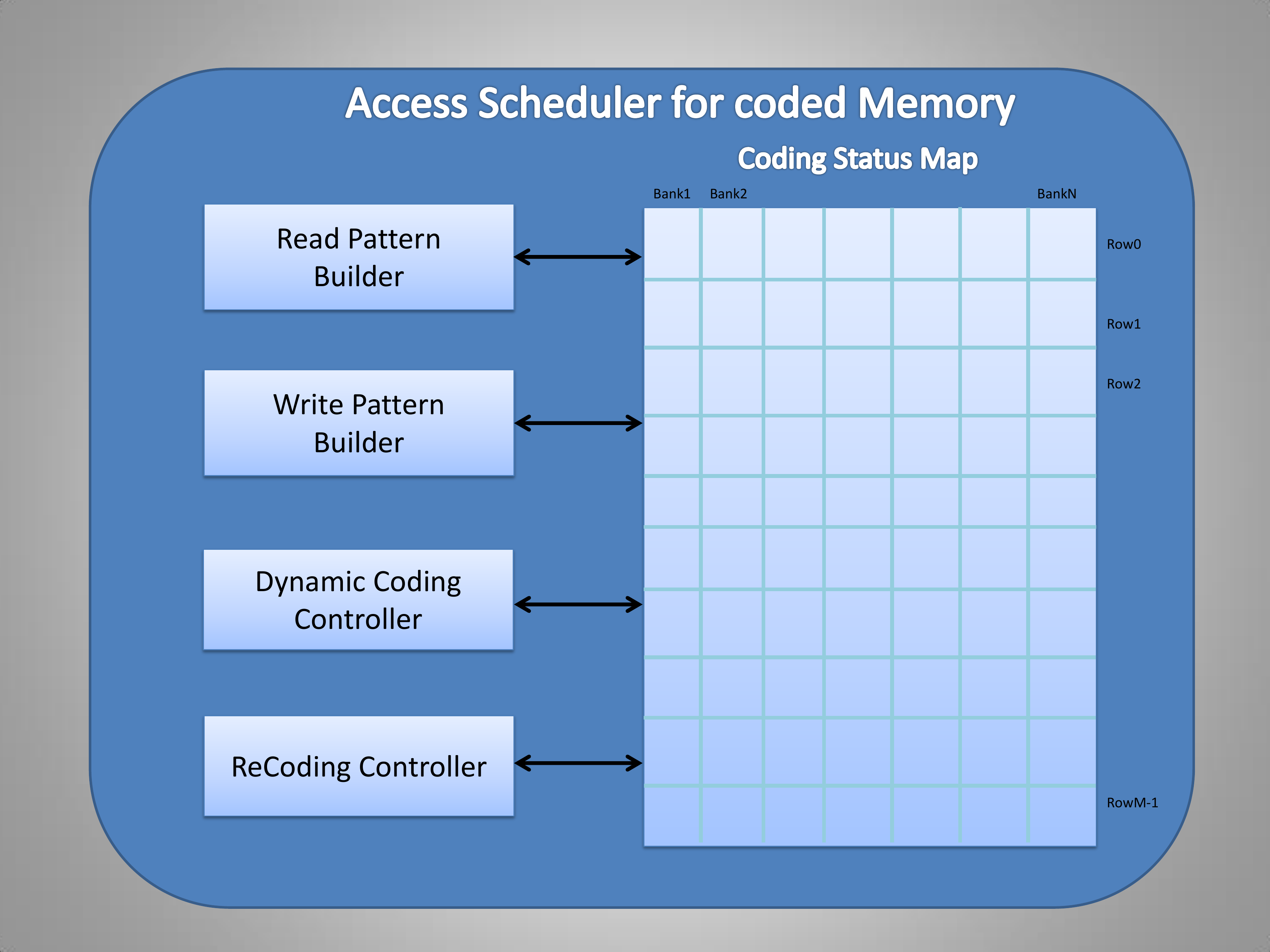}
\caption{
{\it{Pictured here is an illustration of the access scheduler.} }}
\label{fig:coded_access_scheduler}
\end{figure}
%-------------------------
\subsection{Code Status Table}
\label{sec:codeStatusTable}
The code status table keeps track of the validity of elements stored in the data and parity banks. When a write is served to a row in a data bank, any parity bank which is constructed from the data bank will contain invalid data in its corresponding row. Similarly, when the access scheduler serves a write to a parity bank, both the data bank which contains the memory address specified by the write request and any parity banks which utilize that data bank will contain invalid data.

Figure~\ref{fig:coded_access_scheduler} depicts our implementation of the code status table. It contains an entry for every row in each data bank, which can take one of three values indicating 1) the data in both the data bank and parity banks is fresh, 2) the data bank contains the most recent data, or 3) one of the parity banks contains the most recent data. 

\begin{figure}[t!]
	\center{\includegraphics[width=0.56\linewidth]{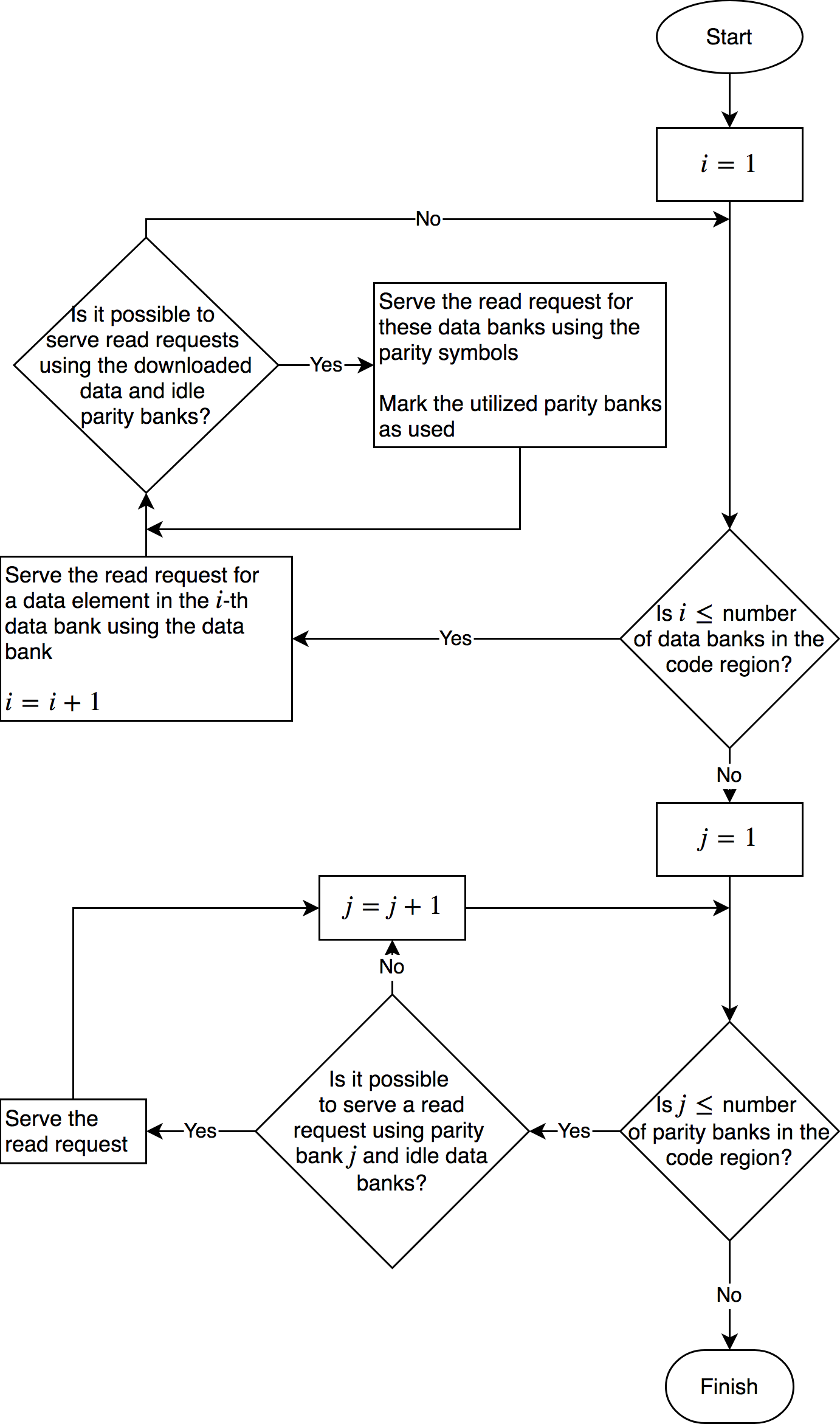}}
	\caption{{\it{Pictured here is a flowchart of the read pattern builder.}}}
	\label{fig:readAlgo}
\end{figure}
%-------------------------
%-----------------------
\begin{figure}[htbp]
	\centering
	\includegraphics[width=0.86\linewidth]{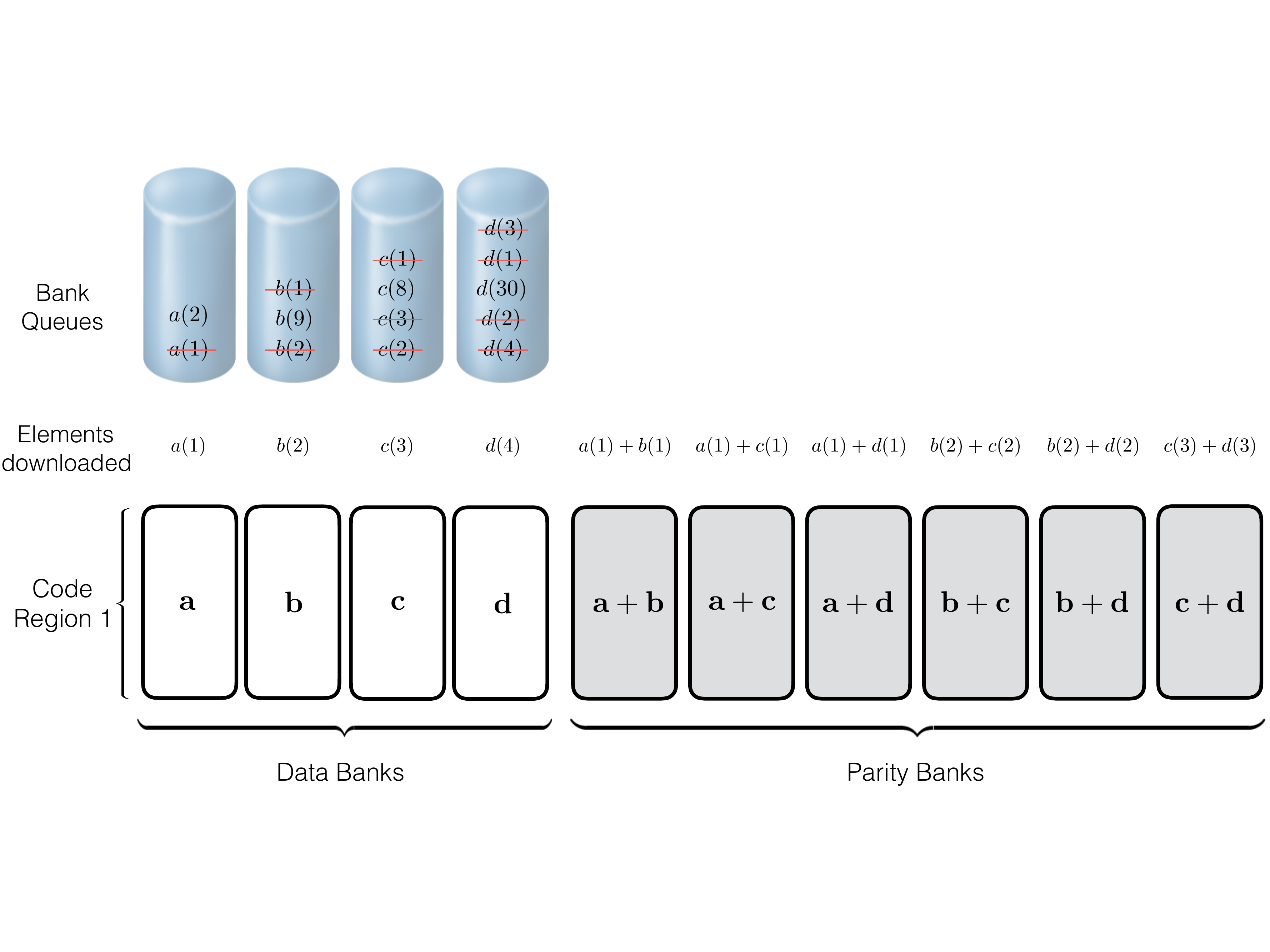}
	\caption{{\it{Illustration of the algorithm to build a read request pattern to be served in a given memory cycle. All the read requests associated with the strike-through elements are scheduled to be served in a given memory cycle. The figure also shows the elements downloaded from all the memory banks in order to serve these read requests.}}}
	\label{fig:readAlgoAccessPattern}
\end{figure}
%------------------------
\subsection{Read pattern builder}
\label{sec:readCodingAlgo}

The access scheduler uses the read pattern builder algorithm to determine which requests to serve using parity banks and which to serve with data banks. The read pattern builder selects which memory requests to serve and determines how requests served by parity banks will be decoded. The algorithm is designed to serve many read requests in a single memory cycle. Figure~\ref{fig:readAlgo} shows our implementation of the read pattern builder.

Figure~\ref{fig:readAlgoAccessPattern} shows an example read pattern constructed by our algorithm. The provided scenario is an example where the parity banks are used to their best effect, because each parity bank is used to serve an additional read request.

\subsection{Write pattern builder}
\label{sec:writeCodingAlgo}
\begin{figure}[htbp]
\centering
	\center{\includegraphics[width=0.66\linewidth]{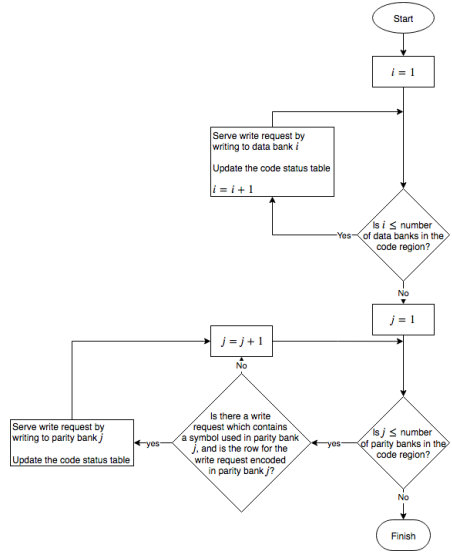}}
	\caption{\it{Pictured here is a flowchart of the write pattern builder.}}
	\label{fig:writeFlow}
\end{figure}

Parity banks allow the memory controller to serve additional write requests per cycle. When multiple writes target a single bank, it can commit some of them to parity banks. The access scheduler implements a write pattern builder algorithm to determine which write requests to schedule in a single memory cycle. Figure~\ref{fig:writeFlow} illustrates our implementation of the write pattern builder. Only when the write bank queues are nearly full does the access scheduler execute the write pattern builder algorithm. 

%-----------------------
%-------------------------

\begin{figure}[t!]
\centering
        \center{ \includegraphics[width=.8\linewidth]{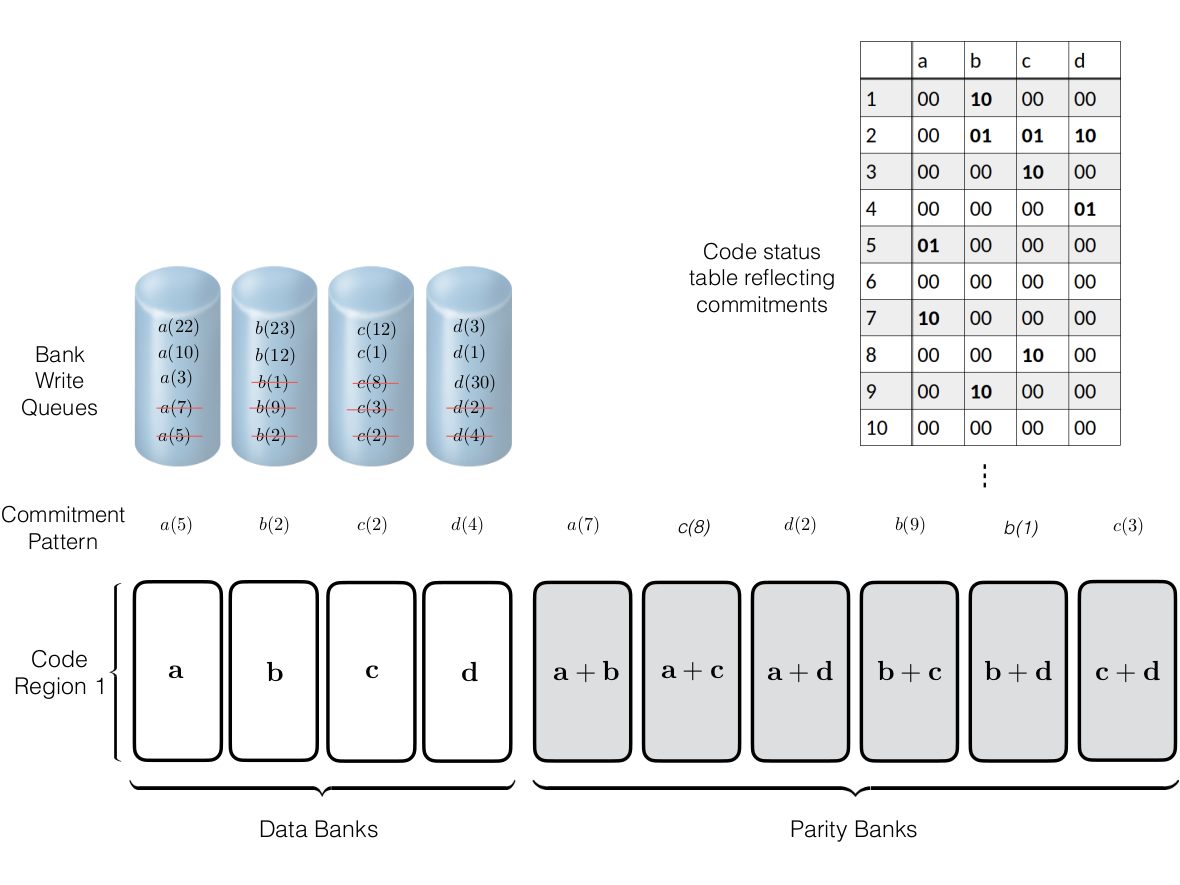}}
	\caption{\it{The behavior of the write pattern builder on a 4-bank memory system is demonstrated here.}}
	\label{fig:writeAlgoAccessPattern}
\end{figure}

Figure~\ref{fig:writeAlgoAccessPattern} shows an example write pattern produced by our algorithm. Parity banks increase the maximum number of write requests from $4$ to $10$. Note that an element which is addressed to row $n$ in a data bank can only be written to the corresponding row $n$ in the parity banks. In this scenario, the write queues for every data bank are full. The controller takes $2$ write requests from each queue and schedules one to the queue's target data bank and the other to a parity bank. The controller also updates the code status table.

Figure~\ref{fig:writeAlgoAccessPattern} also demonstrates how the code status table changes to reflect the freshness of the elements in the data and parity banks. Here, the 00 status indicates that all elements are updated. The 01 status indicates that the data banks contain fresh elements and the elements in the parity banks must be recoded. The 10 status indicates that the parity banks contain fresh elements, and that both data banks and parity banks must be updated.

%-----------------------
%-------------------------
\subsection{ReCoding unit}
\label{sec:recoding}
After a write request has been served, the stale data in the parity (or data) banks must be replaced. The \textit{ReCoding Unit} accomplishes this with a queue of {\em recoding requests}. Every time a write is served, recoding requests are pushed onto the queue indicating which data and parity banks contain stale elements, as well as the bank which generated the recoding request. Requests also contain the current cycle number so that the ReCoding Unit may prioritize older requests.

\subsection{Dynamic Coding}
\label{sec:dynamicCoding}
To reduce memory overhead $\alpha$, parity banks are designed to be smaller than data banks. The dynamic coding block maintains codes for the most heavily accessed memory subregions, so that parity banks are utilized more often.

The {\em dynamic coding} block partitions each memory bank into $\lceil\frac{1}{r}\rceil$ regions. The block can select up to $\frac{\alpha}{r} - 1$ regions to be encoded in the parity banks. A single region is reserved to allow encoding of a new region.

Every $T$ cycles, the dynamic coding unit chooses the $\frac{\alpha}{r} - 1$ regions with the greatest number of memory accesses. The dynamic coding unit will then encode these regions in the parity banks. If all the selected regions are already encoded, the unit does nothing. Otherwise, the unit begins encoding the most accessed region. Once the dynamic coding unit is finished encoding a new region, the region becomes available for use by the rest of the memory controller. A memory region of length $r$ is reserved by the dynamic coding unit for constructing new encoded regions, and a memory region of length $\alpha - r$ is reserved for active encoded regions. If the memory ceiling $\alpha - r$ is reached when a new memory region is encoded, the unit evicts the least frequently used encoded region.

%%%%%%%%%%%%%%%%%%%%%%%%%%%%%%%%%%%%%%%%%%%%%%%%%%%%%%%%
% Experimental Methodology
%%%%%%%%%%%%%%%%%%%%%%%%%%%%%%%%%%%%%%%%%%%%
\section{Experiments}
\label{sec:experimentalmethodology}

In this section, we discuss our method for evaluating the performance of the proposed memory system. We utilize the PARSEC v2.1 and v3.0 benchmark suites with the gem5 simulator~\cite{bienia09parsec2, parsec_2_1_m5} to generate memory traces. Next, we run the Ramulator DRAM simulator~\cite{Ramulator} to measure the performance of the proposed memory system. Next, we compare the baseline performance of the Ramulator DRAM simulator against a modified version which implements the proposed system.

\subsection{Memory Trace Generation}
The PARSEC benchmark suite was developed for chip multiprocessors and is composed of a diverse set of multithreaded applications~\cite{bienia09parsec2}. These benchmarks allow us to observe how the proposed memory system performs in dense memory access scenarios. 

The gem5 simulator~\cite{parsec_2_1_m5} allows us to select the processor configuration used when generating the memory traces. We use $8$ processors in all simulations. The PARSEC traces can be split into regions based on computation type. We focus on regions which feature parallel processing because they have the greatest likelihood of bank conflicts.

Many attributes affect the performance of our proposed memory system, most importantly the density of traces, the overlap of memory accesses among processors, and the stationarity of heavily utilized memory regions.

\begin{figure}[htbp]
		\center{\includegraphics[width=.8\linewidth]{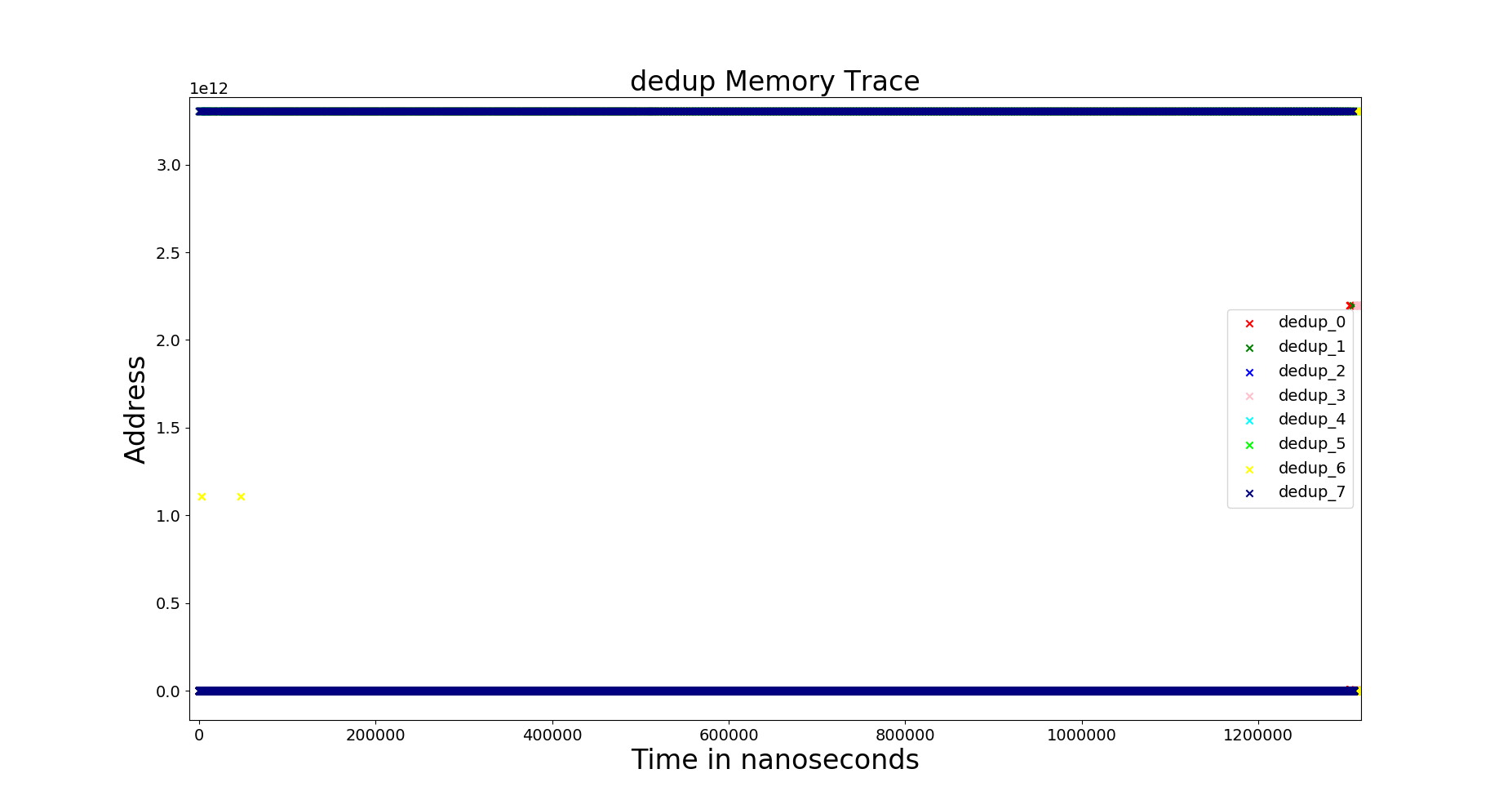}}
		\caption{\it{The memory access pattern of the dedup PARSEC benchmark.}}
		\label{fig:dedup_benchmark}
\end{figure}

We find that memory access patterns that occupy consistent bands of sequential memory addresses benefit most from our proposed memory system. Figure~\ref{fig:dedup_benchmark} shows the access pattern of one of the dedup PARSEC benchmark.

\begin{figure}[htbp]
		\center{\includegraphics[width=.8\linewidth]{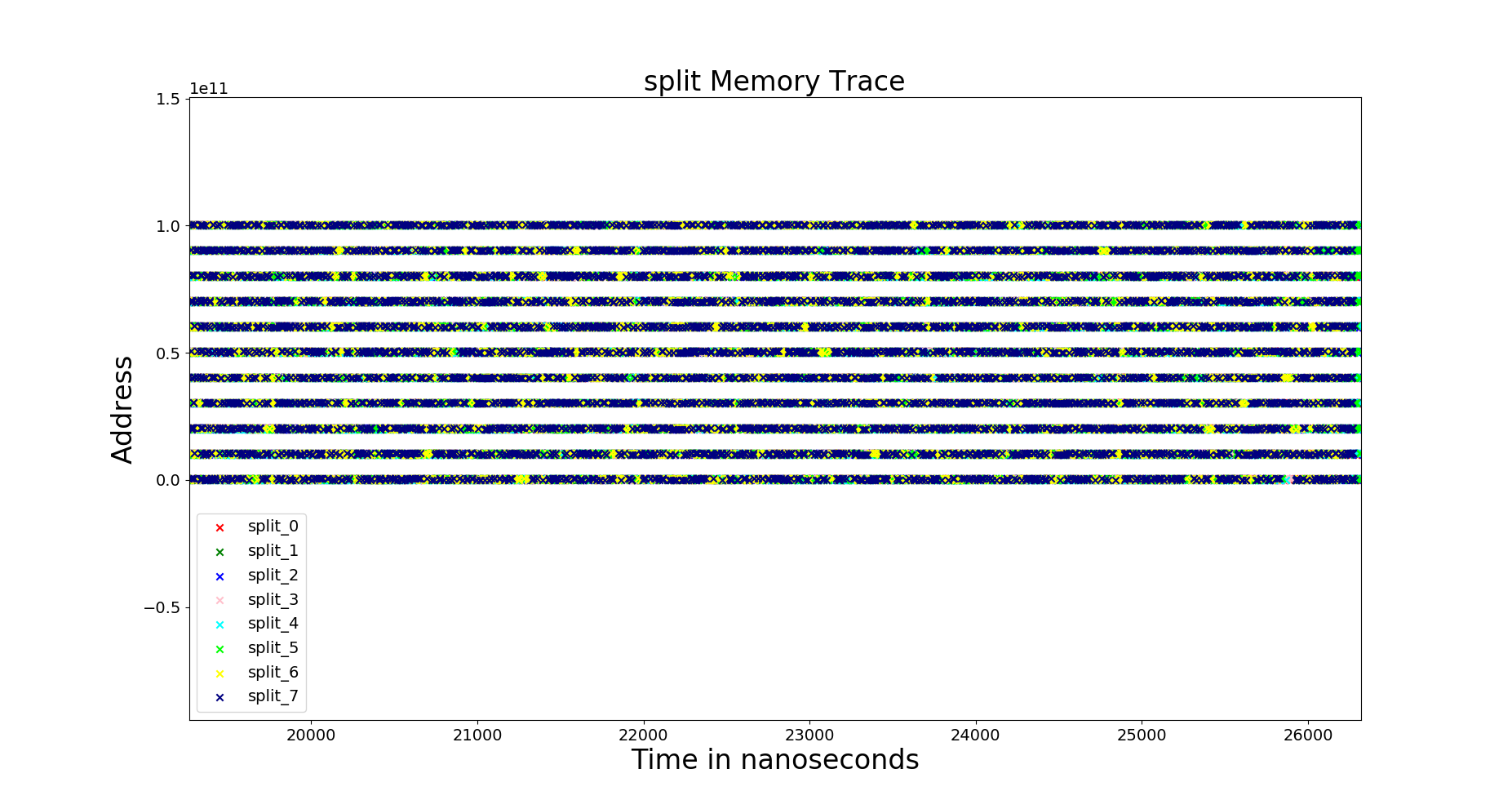}}
		\caption{\it{The vips benchmark after splitting the primary access bands into multiple additional bands.}}
		\label{fig:vips_split}
\end{figure}

\begin{figure}[htbp]
		\center{\includegraphics[width=.8\linewidth]{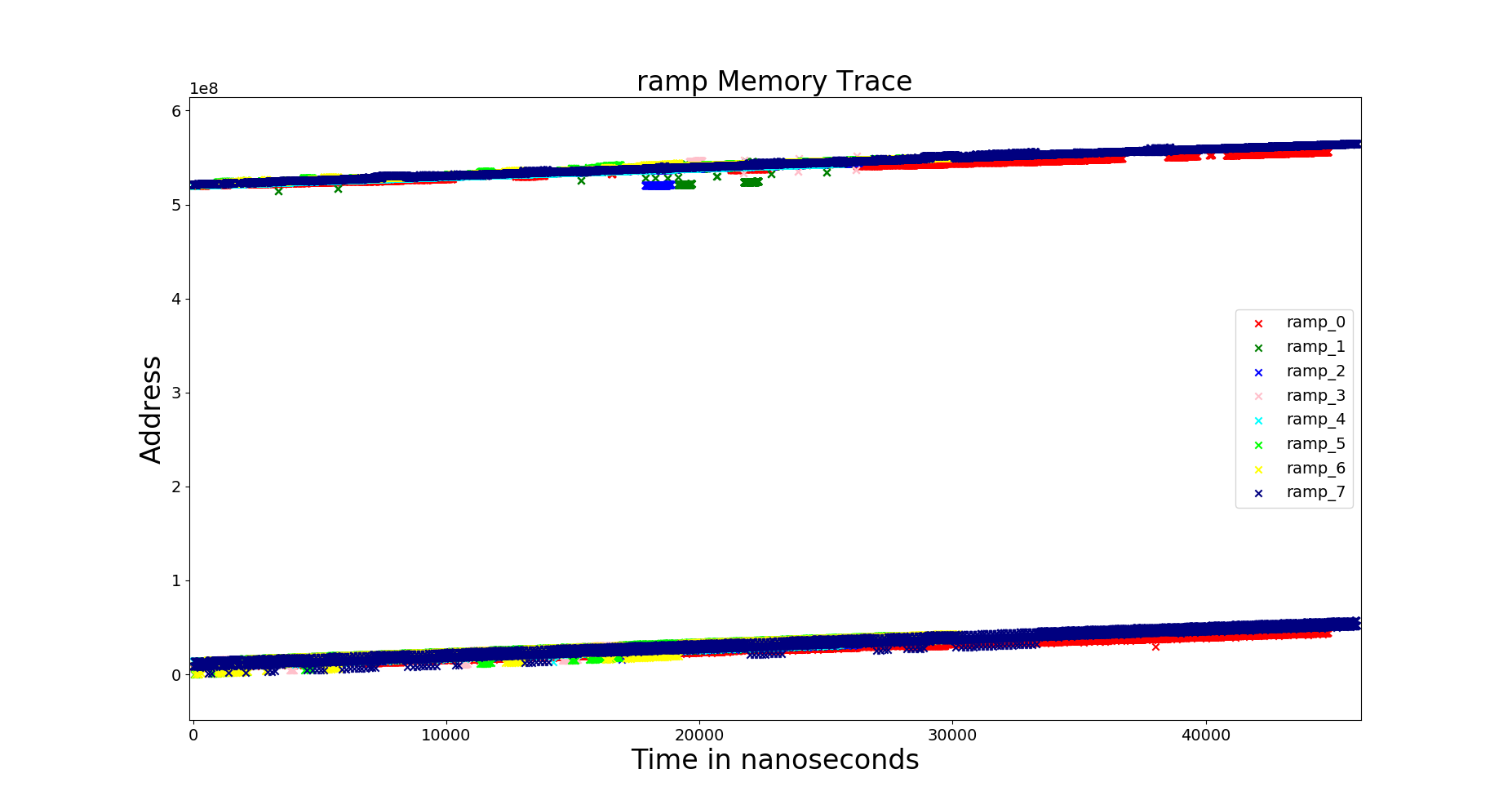}}
		\caption{\it{The vips benchmark after adding a ramp to the major memory bands.}}
		\label{fig:vips_ramp}
\end{figure}

We augment the PARSEC benchmarks in two ways to test our system in additional scenarios, shown in Figures~\ref{fig:vips_split} and~\ref{fig:vips_ramp}, respectively. First we split the given memory bands to simulate an increased number of bands in the system. Next, we introduce dynamic memory access patterns by adding a linear ramp to the previously static address locations. 

\subsection{Ramulator}

We use the Ramulator DRAM simulator to compare the number of CPU cycles required to execute the PARSEC memory traces. First, we use the original Ramulator simulator to measure a baseline number of CPU cycles. Then we implement the proposed memory system and use the modified Ramulator (fixing all other configuration) to calculate improvements over the baseline. Our simulations vary the overhead parameter $\alpha$.

%%%%%%%%%%%%%%%%%%%%%%%%%%%%%%%%%%%%%%%%%%%%%%%%%%%%%%%%
% Results
%%%%%%%%%%%%%%%%%%%%%%%%%%%%%%%%%%%%%%%%%%%%
\subsection{Simulation Results}
\label{sec:simulation}

Given sufficient memory overhead, we see a consistent 25\% reduction in CPU cycles over the baseline simulation, with Coding Scheme I generally performing best. 

\begin{figure}[htbp]
		\center{\includegraphics[width=.80\linewidth]{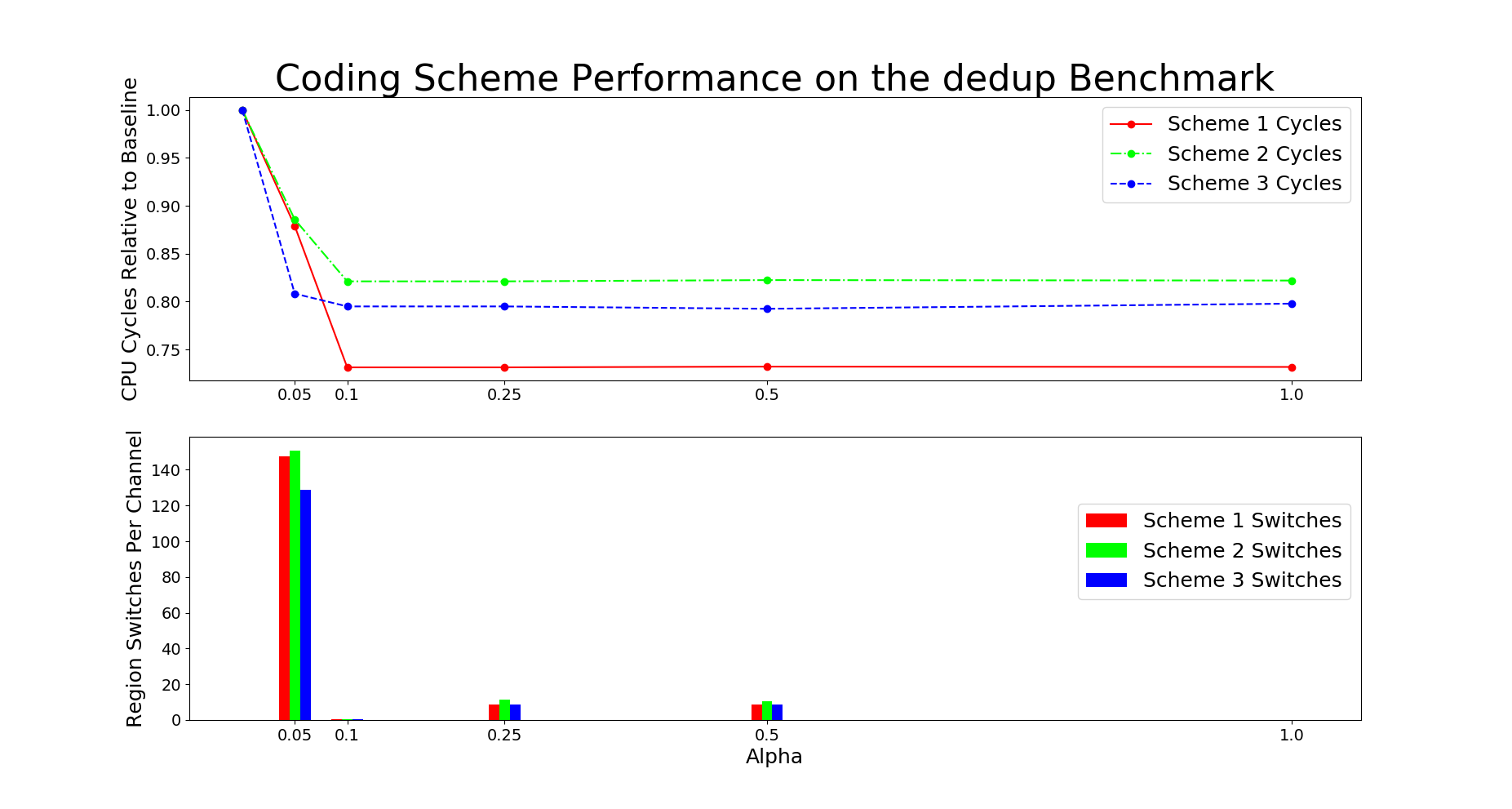}}
		\caption{\it{The simulation results for the dedup PARSEC benchmark. The line plot represents the number of CPU cycles executed and the bar plot represents the number of times the dynamic coding unit chooses to encode a new memory region. }}
		\label{fig:dedup_results}
\end{figure}

The proposed memory system performs consistently across the PARSEC benchmarks, and the three proposed schemes yield similar results. Figure~\ref{fig:dedup_results} shows the simulation results for the dedup benchmark with a memory partition coefficient $r = 0.05$. The plot shows that the number of CPU cycles is reduced by $73\%-83\%$ once sufficient memory overhead $\alpha$ is used. 

We also see that the number of memory region switches performed by the dynamic encoder. When $\alpha = 1$, the number of switches is always zero as expected because the dynamic encoder never needs to switch regions. The performance remains consistent for $\alpha > 0.1$. With this amount of overhead, the memory system finds and encodes the two heavily accessed memory bands in each of the PARSEC benchmarks. This is because $\lfloor\frac{\alpha}{r}\rfloor = 2$, which means we can select $2$ regions to encode. 

When $\alpha = 0.05$, the number of coded region switches is very high because the memory system vacillates between the two most heavily accessed bands. When $\alpha = .1$, both of them can be encoded. We see a small numbers of switches when $\alpha \geq 0.25$ because the memory system is encoding less heavily accessed memory bands with little impact on number of CPU cycles.

\subsubsection{Augmented PARSEC}

Results on the the augmented PARSEC traces show that our system improves over the baseline to a lesser extent.

\begin{figure}[htbp]
		\center{\includegraphics[width=.80\linewidth]{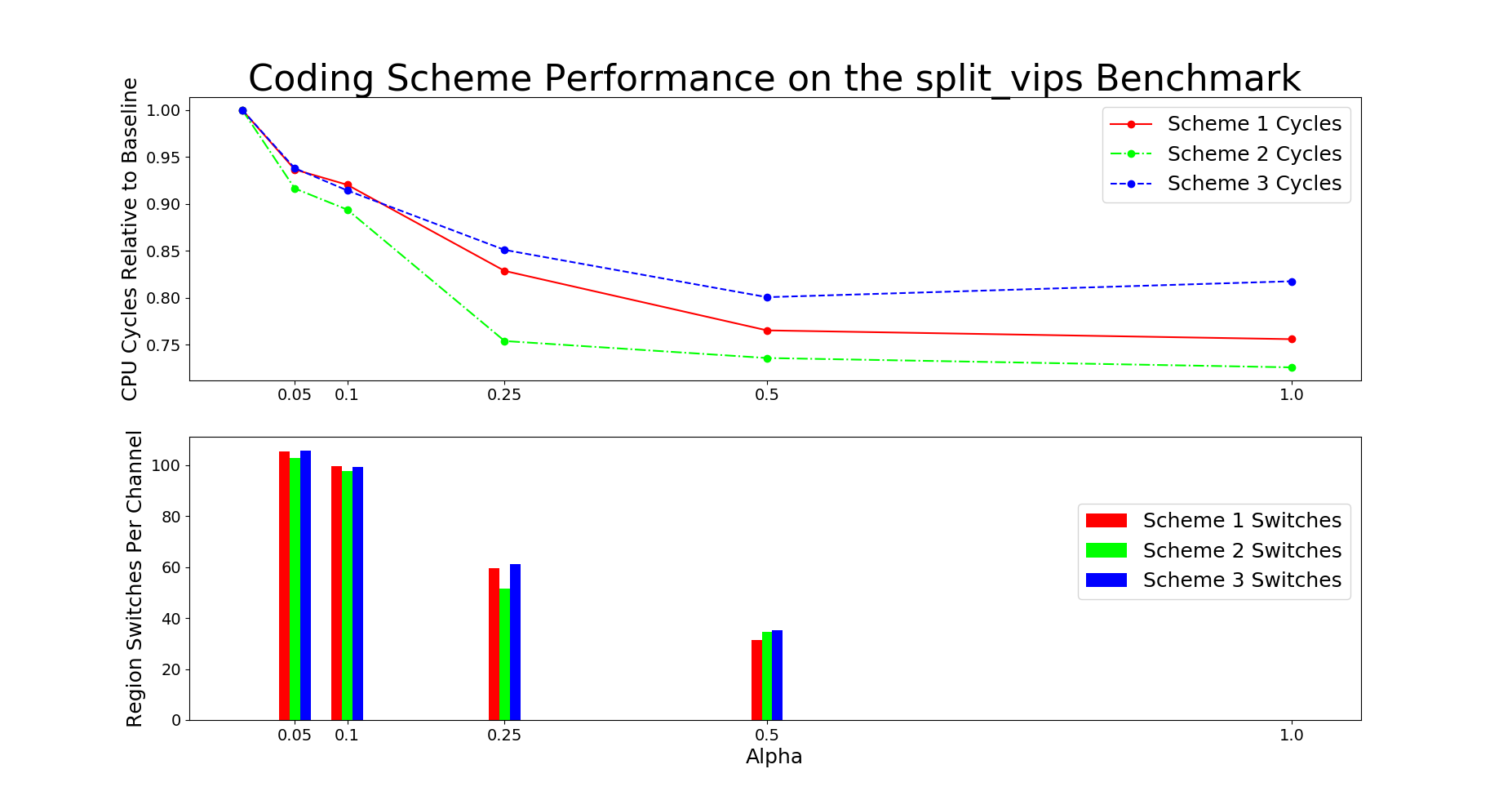}}
		\caption{\it{The simulation results of the augmented vips trace pictured in Figure~\ref{fig:vips_split}.}}
		\label{fig:vips_split_result}
\end{figure}

Figure~\ref{fig:vips_split_result} shows that for a large number of memory bands, we can achieve the same performance as before only by increasing the memory overhead or increasing the memory partition coefficient.

\begin{figure}[htbp]
		\center{\includegraphics[width=.80\linewidth]{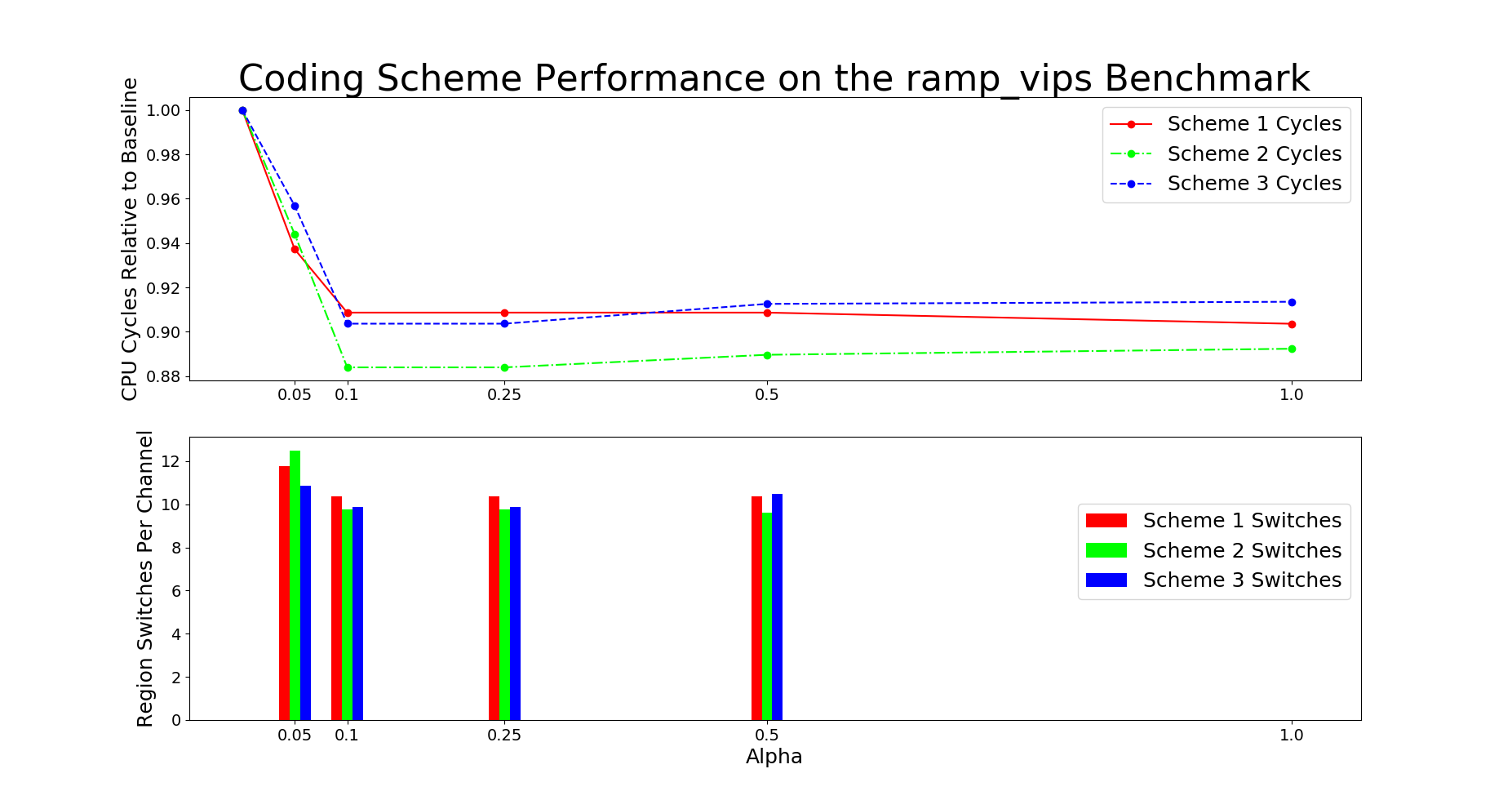}}
		\caption{\it{The simulation results of the augmented vips trace pictured in Figure~\ref{fig:vips_ramp}.}}
		\label{fig:vips_ramp_result}
\end{figure}

Figure~\ref{fig:vips_ramp_result} shows the results of the ramp augmentation. Here we see that our system struggles to adapt to a constantly changing primary access region. 

%%%%%%%%%%%%%%%%%%%%%%%%%%%%%%%%%%%%%%%%%%%%%%%%%%%%%%%%
% Conclusion
%%%%%%%%%%%%%%%%%%%%%%%%%%%%%%%%%%%%%%%%%%%%

%%%%%%%%%%%%%%%%%%%%%%%%%%%%%%%%%%%%%%%%%%%%%%
% Acknowledgements
%%%%%%%%%%%%%%%%%%%%%%%%%%
\section{Conclusion}

Our proposed design emulates multi-port memory using coding techniques with single-port memory, and it is able to speed up the execution time of the PARSEC benchmarks. We are able to support multiple read and writes in a single memory cycle, and compared to replication-based methods we use far less memory and therefore power and chip area. However, The design of a parity storage requires additional logic at the memory controller to encode/decode and schedule read/writes. Further iterations on our design may include using idle banks to prefetch symbols and improvements to the read and write pattern builders.

\section{Acknowledgements}
This document is derived from previous conferences, in particular HPCA 2017.  We thank Daniel A. Jimenez,  Elvira Teran for their inputs.

%%%%%%%%% -- BIB STYLE AND FILE -- %%%%%%%%
\bibliographystyle{IEEEtran}
\bibliography{references}
%%%%%%%%%%%%%%%%%%%%%%%%%%%%%%%%%%%%

\newpage 
\end{document}